\documentclass[lettersize,journal,twoside]{IEEEtran}
\usepackage{cite}
\usepackage[cmex10]{amsmath}
\usepackage{amssymb,amsfonts}
\usepackage{stfloats}
\usepackage{amsthm}
\usepackage{float}
\usepackage{graphicx}
\usepackage{textcomp}
\usepackage{xcolor}
\usepackage[tight,footnotesize]{subfigure}
\newcommand{\mr}{\mathrm}
\usepackage{url}
\newcommand{\bb}{{\rm BB}}
\newcommand{\rf}{{\rm RF}}
\newcommand{\tabincell}[2]{\begin{tabular}{@{}#1@{}}#2\end{tabular}}
\usepackage{tablefootnote}
\usepackage{threeparttable}
\normalsize

\floatstyle{ruled}
\newfloat{algorithm}{tbp}{loa}
\providecommand{\algorithmname}{Algorithm}
\floatname{algorithm}{\protect\algorithmname}

\theoremstyle{plain}

\theoremstyle{definition}

\theoremstyle{plain}

\theoremstyle{plain}

\theoremstyle{plain}

\usepackage{tikz}
\usepackage{marvosym}
\usepackage{pgf}\usetikzlibrary{calc,arrows}
\usepackage{algorithm}
\usepackage{algpseudocode}

\providecommand{\lemmaname}{Lemma}
\providecommand{\theoremname}{Theorem}

\makeatother

\providecommand{\corollaryname}{Corollary}
\providecommand{\definitionname}{Definition}
\providecommand{\lemmaname}{Lemma}
\providecommand{\propositionname}{Proposition}
\providecommand{\theoremname}{Theorem}
\ifCLASSINFOpdf
\else
\fi

\begin{document}
\title{Design and Analysis of Wideband In-Band-Full-Duplex FR2-IAB Networks}
\author{Junkai~Zhang,~\IEEEmembership{Student~Member,~IEEE}, Haifeng Luo, Navneet~Garg,~\IEEEmembership{Member,~IEEE,} Abhijeet~Bishnu,~\IEEEmembership{Member,~IEEE,} Mark~Holm,~\IEEEmembership{Member,~IEEE,}
and~Tharmalingam~Ratnarajah,~\IEEEmembership{Senior~Member,~IEEE}
\thanks{Manuscript received March 30, 2021; revised August 26, 2021, and November 3, 2021; accepted November 4, 2021. The work of J. Zhang, H. Luo, N. Garg, and A. Bishnu was supported by
the research grant from Huawei Technologies (Sweden)
AB. The work of T. Ratnarajah was supported by the U.K. Engineering and Physical Sciences Research
Council (EPSRC) under Grant EP/P009549/1. The associate editor coordinating the review of this article and approving it for publication was Trung Q. Duong. \textit{(Corresponding author: Junkai Zhang.)}}%
\thanks{J. Zhang, H. Luo, N. Garg, A. Bishnu and T. Ratnarajah are with Institute for Digital Communications, The University of Edinburgh, Edinburgh, EH9 3FG, UK (e-mail: \{jzhang15, hluo2, ngarg, abishnu, T. Ratnarajah\}@ed.ac.uk.)}%
\thanks{M. Holm is with Radio Basestation Systems Department, Huawei Technologies (Sweden) AB, Gothenburg, Sweden. (e-mail: mark.holm@huawei.com).}
}

\markboth{IEEE Transactions on Wireless Communications,~Vol.~x, No.~x, xx~2021}%
{Zhang \MakeLowercase{\textit{et al.}}: Design and Analysis of Wideband In-Band-Full-Duplex FR2-IAB Networks}


\maketitle
\begin{abstract}
This paper develops a 3GPP-inspired design for the in-band-full-duplex (IBFD) integrated access and backhaul (IAB) networks in the frequency range 2 (FR2) band, which can enhance the spectral efficiency (SE) and coverage while reducing the latency. However, the self-interference (SI), which is usually more than 100 dB higher than the signal-of-interest, becomes the major bottleneck in developing these IBFD networks. We design and analyze a subarray-based hybrid beamforming IBFD-IAB system with the RF beamformers obtained via RF codebooks given by a modified Linde-Buzo-Gray (LBG) algorithm. The SI is canceled in three stages, where the first stage of antenna isolation is assumed to be successfully deployed. The second stage consists of the optical domain (OD)-based RF cancellation, where cancelers are connected with the RF chain pairs. The third stage is comprised of the digital cancellation via successive interference cancellation followed by minimum mean-squared error baseband receiver. Multiuser interference in the access link is canceled by zero-forcing at the IAB-node transmitter. Simulations show that under 400 MHz bandwidth, our proposed OD-based RF cancellation can achieve around 25 dB of cancellation with 100 taps. Moreover, the higher the hardware impairment and channel estimation error, the worse digital cancellation ability we can obtain.
\end{abstract}


\IEEEpeerreviewmaketitle

\section{Introduction}\label{I}
\IEEEPARstart{F}{requency} range 2 (FR2) band (i.e., millimeter wave) communications have been identified as the key technology for the  beyond  fifth-generation (5G) wireless communications to provide much larger bandwidth, narrower beam, and high data rate services. Different from the FR1 band ($\leq 7.225$ GHz), in the FR2 band ($\geq 24.250$ GHz), high path loss and blockages become the major obstacles for broader coverage. However, the short wavelengths at the FR2 frequencies facilitate the deployment of large-scale antenna arrays, which could compensate for such high losses with highly directional narrow beamforming and provide reliable transmission quality \cite{8594596,7448873}. 

In the recent 3rd Generation Partnership Project (3GPP) technical report TR 38.874 (Rel. 16) \cite{3gpp}, the integrated access and backhaul (IAB) networks have been proposed for the FR2 band communications, where only IAB donors connect with the core network by fiber. IAB-nodes can wirelessly communicate with both the access and the backhaul links as well as perform IAB-specific tasks such as resource allocation, route selection, and optimization \cite{MAG}. This novel architecture enables cheap and dense deployment while extending the coverage in FR2 bands. Despite the visible advantages of this architecture, the study of IAB networks is still in its infancy.

In-band-full-duplex (IBFD) transmission, which has been treated as another breakthrough for beyond 5G wireless communications, breaks the rule that downlink and uplink communications should occur in different time/frequency slots. In the IAB networks, IAB-nodes are preferred to run under the IBFD mode \cite{tong}. Compared with the half-duplex (HD), thanks to simultaneous transmissions, the IBFD mode can almost double the spectral efficiency (SE) without the need for the large guard time/band arranged in standard time-division duplex and frequency-division duplex systems \cite{8246856,7815376}. However, the major obstacle of IBFD communications is the existence of strong self-interference (SI), which is usually seen as more than 100 dB stronger than the signal of interest \cite{7224732}. Therefore, finding efficient SI cancellation (SIC) techniques is important for IBFD operation and has recently been a popular research topic. Through hardware prototype measurements in 28 GHz, authors in \cite{suk2021duplex} evaluate the framework’s link-level SI reduction in the propagation domain and system-level performance to verify the feasibility of IBFD-IAB systems; however, the large-scale antenna array and hybrid precoding were not considered.

For the wideband IBFD-FR2 communications, we propose a three-stage SIC, which consists of the antenna isolation stage (i.e., by isolating the transceiver antennas electromagnetically for passive cancellation) \cite{6702851}, the analog cancellation (A-SIC) stage (i.e., by establishing a circuit canceler between each transceiver pair to replicate the SI channel as accurately as possible) \cite{8171089}, and the digital cancellation (D-SIC) stage (i.e., handling of the residual SI (RSI) left by previous stages by designing efficient beamformers) \cite{7224732,5757643,9431171}. In the A-SIC, the conventional micro-strip analog canceler requires a huge number of taps for wideband SIC. However, due to the large insertion losses and realization of hundreds of taps, wideband SIC becomes infeasible in practice. Besides, it is challenging for the micro-strip analog canceler to be directly extended to FR2 band scenarios due to the hardware limitation (i.e., RF components usually do not have such processing properties at the FR2 band). Thus, a hardware efficient optical domain (OD)-based analog canceler has been investigated in \cite{8796422} for the single antenna system. However, OD-based A-SIC for multi-antenna systems or IAB networks is lacking in the literature.

Due to the use of large-scale array systems, the traditional full digital beamforming scheme for the FR1 band becomes expensive to implement for the FR2 band. Thus, towards the need for cost-friendly system design, hybrid beamforming has become a powerful and economical tool in large-scale array systems, which reduces the requirement on the number of RF chains and simplifies the system complexity \cite{8644218}. Based on the extension of the standard Orthogonal Matching Pursuit (OMP) algorithm, a novel hybrid beamforming design was proposed in \cite{7827111}. Compared with the fully connected hybrid beamforming structure \cite{7448873}, to improve the deployment cost and guarantee the similar performance of the system, authors in \cite{8594596,7880698}, and \cite{8295113} develop a subarray hybrid beamforming structure, where one RF chain only connects with a portion of antenna arrays. However, the works that consider the wideband IBFD multi-user IAB networks with subarray hybrid beamforming in the FR2 band still need more investigation.

The hybrid beamforming design algorithms in \cite{8594596,7448873,8644218,7827111,7880698} need to access the large and sparse channel matrix, which is hard to acquire in reality. Although the compressed sensing-based channel estimation approaches are presented in \cite{7961152}, it is difficult to realize in practical scenarios. Instead, the RF effective channel is estimated using standard estimation methods in practice, where the RF precoding and combining matrices are selected from the pre-defined codebooks. In \cite{7448873}, the RF codebook is designed by the Lloyd type algorithm. A $K$-means-based beam codebook is proposed by Mo \textit{et al.}, whose codewords are defined by maximizing the beamforming gain \cite{8767936}. Unfortunately, their vector-wise codebooks may lead to a low-rank beamforming matrix, which directly amounts to a loss in the degrees of freedom, especially when the number of RF chains is more than one.

Further, the hardware impairments (HWI), which takes into account the imperfection in the hardware, such as oscillators noise, amplifiers noise, non-linearities in the digital-to-analog converters (DACs) and the analog-to-digital converters (ADCs), and etc., have not been considered in most of the studies yet. Authors in \cite{6280258} have mentioned that the independent Gaussian model can optimally capture those combined non-ideal hardware effects.

Based on the above motivations, in this paper, we investigate the design and analysis of multiuser FR2-IBFD-IAB networks with subarray-based hybrid beamforming. The contributions of this work are given as follows:
\begin{itemize}
    \item \textit{RF codebook design and RF effective channel estimation:} For the subarray hybrid beamforming scheme, the RF precoders and combiners are selected by scanning from the matrix-wise codebooks, designed with our modified mean squared error (MSE)-based Linde-Buzo-Gray (LBG) algorithm, and the RF effective channel can be then estimated with standard estimation methods. Simulations show that, with the proposed codebooks, we can achieve a similar SE as that with infinite resolution phase shifters (PSs) without suffering from low rank quantized beamforming matrices.

    \item \textit{Staged SIC:} We propose a staged SIC scheme in this paper, where the A-SIC is realized by the OD-based canceler connected with the RF chain pairs on the IAB-node to reduce the space and cost. Compared with the conventional micro-strip analog canceler, our canceler can provide a significant number of true delay lines for wideband operations and have good frequency-flatness. Simulations show that with our OD-based canceler, 25 dB of A-SIC can be achieved with about 100 taps over 400 MHz bandwidth.
    
    \item \textit{System Analysis with RSI:} In order to explore how the RSI caused by the HWI and RF effective channel uncertainties can affect the performance of the IBFD system, we analyze the SE of the backhaul link by varying the HWI factors and SI RF effective channel estimation errors. Simulation results show that as SNR increases, the system becomes more vulnerable to the RSI; however, the tolerance is improved when increasing the codebook size. It is also shown that at lower RSI values, IBFD operation doubles the SE compared to that of the HD.
\end{itemize}

The rest of the paper is organized as follows. In Section~\ref{Sec2}, the system model and channel models are identified, followed by introducing the OD-based analog canceler design in Section~\ref{asic}. Then, the modified LBG algorithm for the RF codebook design is proposed in Section~\ref{RFest}, where RF effective channels are estimated with selected RF beamformer pairs. Next, D-SIC is processed in Section~\ref{digital}. In Section~\ref{Sec4}, the SE expressions are evaluated, followed by the design of BB beamformers for both backhaul and access links. Finally, some simulation results and a brief conclusion are shown in Section~\ref{sec5} and Section~\ref{sec6}, respectively.

\textit{Notations:} $\mathcal{B}, \mathbf{B}, \mathbf{b}$, $b$ represent a set, a matrix, a vector, and a scalar, respectively. $\mathbf{B}^{H}, \mathbf{B}^{-1}$, and $\mathbf{B}^T$ are the Hermitian, inverse, and transpose of $\mathbf{B}$, respectively. $|\mathcal{B}|$ is the cardinality of $\mathcal{B}$. $\lVert\mathbf{B}\rVert_F$, $|\mathbf{B}|_{mn}$, $\mr{det}\{\mathbf{B}\}$, and $\mr{tr}[\mathbf{B}]$ are the Frobenius norm, absolute value of the ($m,n$)th entry, determinant, and trace of $\mathbf{B}$, respectively. $\lVert\mathbf{b}\rVert_{2}$ is the L2-norm of $\mathbf{b}$. $\lVert b\rVert$ is the norm of $b$. $arg(\mathbf{B})$ takes the angle of each entry of $\mathbf{B}$. $\mr{diag}[\mathbf{B}]$ takes the diagonal elements of the matrix. $\mr{blkdiag}[\mathbf{B}_1,\mathbf{B}_2]$ is the block diagonal matrix formed by matrix $\mathbf{B}_1$ and $\mathbf{B}_2$.  $[\mathbf{B}]_{:,1:n}$ and $[\mathbf{B}]_{m,n}$ denote the first $n$ columns and the $(m,n)$th entry of $\mathbf{B}$, respectively. $\mr{Cov}[\mathbf{B}]$ is the covariance matrix, i.e., $\mathbb{E}\{\mathbf{B}\mathbf{B}^H\}$.  $\odot$ indicates the  Hadamard product. $d(\cdot,\cdot)$ is the distance measurement. $\mathcal{CN}(m,n)$ denotes a complex Gaussian distribution with mean value of $m$ and variance $n$, and $\mathbf{I}_K$ is the $K\times K$ identity matrix.

\section{System and Channel Models}\label{Sec2}
\subsection{System Model}\label{II}
\begin{figure}[!t]
\centering
\includegraphics[width=0.5\textwidth]{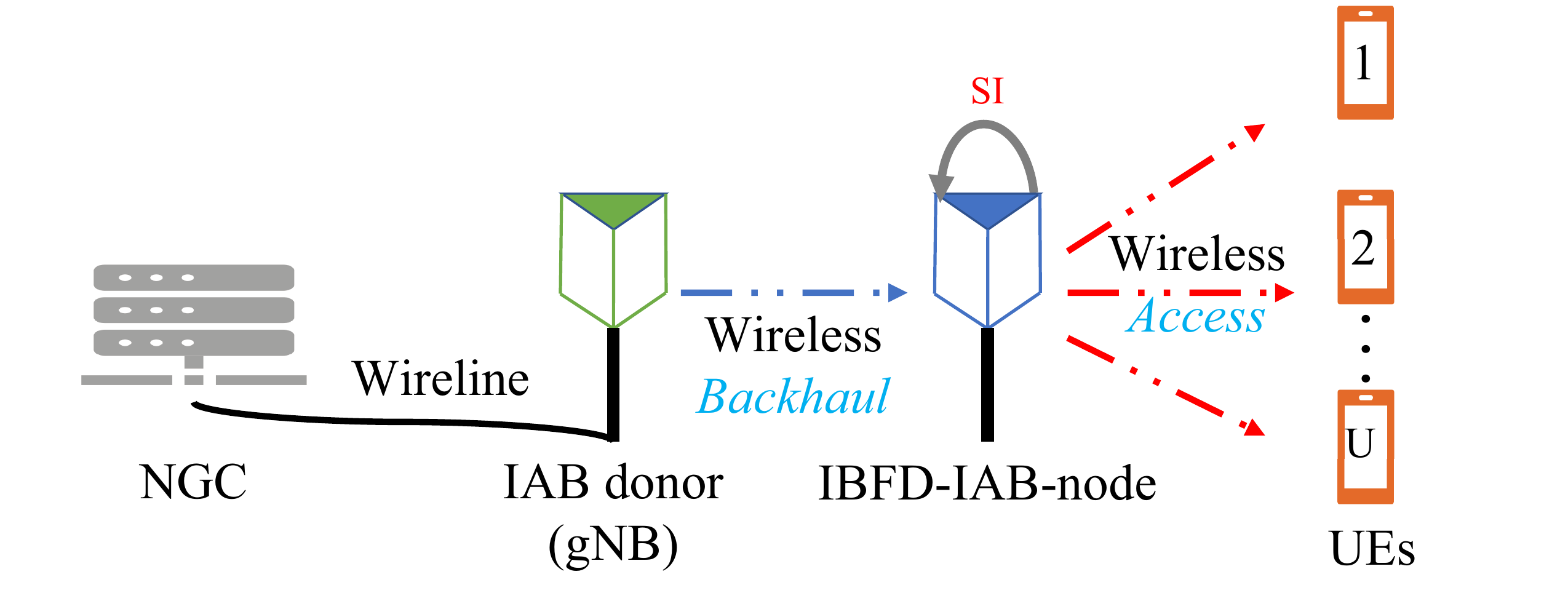}
\caption{Illustration of a single-cell IBFD-IAB multiuser network under the SA deployment.}
\label{SA}
\end{figure}

In this subsection, the system model is described for the wideband FR2-IBFD-IAB multiuser networks. According to the technical specifications--TR 38.874 (Rel. 16) provided by the 3GPP, standalone (SA) and non-standalone (NSA) are two typical deployments considered for IAB networks \cite{3gpp}. In this work, we consider the downlink of a single-cell FR2-IBFD-IAB multiuser network with SA deployment\footnote{The reason why SA structure is considered in this work is that the NSA architecture permits IAB-nodes and UEs to communicate with both 4G base stations (i.e., eNBs) as well as 5G base stations (i.e., gNBs); however, SA only allows connections with 5G base stations, which is considered for future wireless communication network environment. With minor modifications, the present design and analysis can be used for NSA as well.}, which consists of the following parts, that are
\begin{itemize}
    \item an IAB donor, also called gNB, which is a single logical node and acts as the base station;
    \item an IBFD-IAB-node, which contributes SI from its transmitter to its receiver;
    \item $U$ downlink user-equipments (UEs).
\end{itemize}
The IAB donor connects to the 5G next-generation core (NGC) network by fiber and communicates with the IAB-node through a wireless backhaul link. The IAB-node serves the users by wireless access links. Note that, in this work, the IAB donor only provides backhaul link service. An illustration of this IBFD-IAB multiuser network used in this work is depicted in Fig.~\ref{SA}, and more information about the 3GPP architecture can be found in our recent work \cite{MAG}.

\begin{figure*}[!t]
\centering
\includegraphics[width=\textwidth]{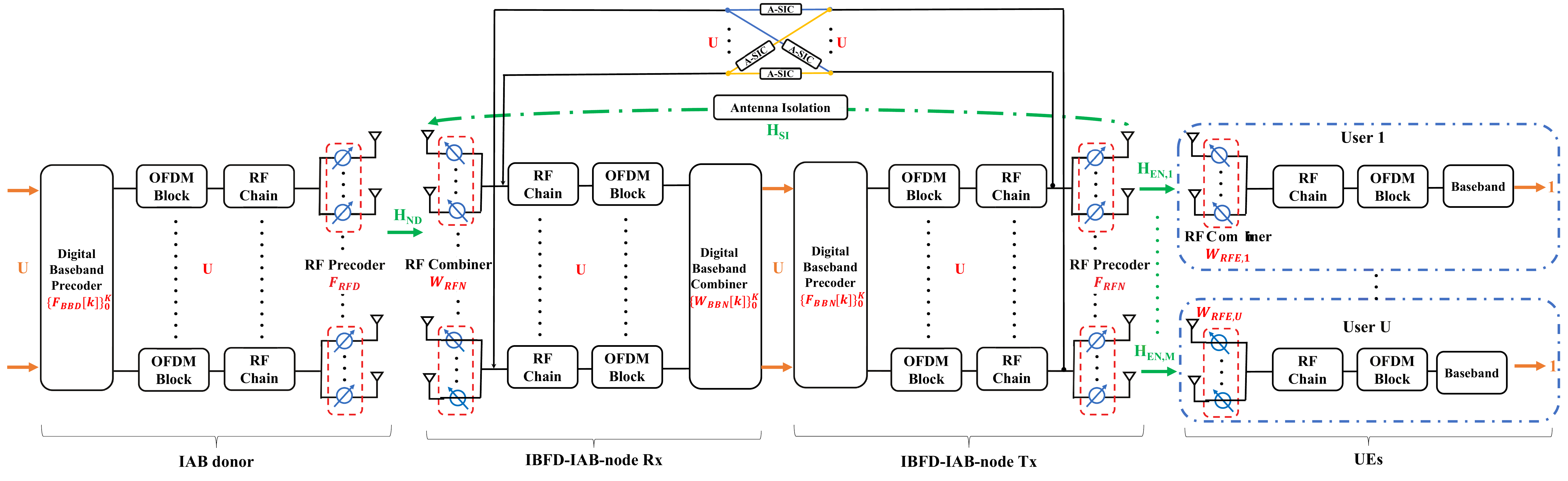}
\caption{Illustration of a wideband FR2-IBFD-IAB multiuser system with subarray hybrid beamforming.}
\label{sys}
\end{figure*}

The IAB donor and IAB-node are equipped with the subarray-based hybrid beamforming structure \cite{7880698}, where each RF chain only connects with a portion of antenna elements. Compared with the fully connected structure \cite{7880698}, the subarray structure provides a cost-efficient solution for connecting RF chains to the antenna arrays. The number of subarrays (RF chains) at the IAB-donor and IAB-node is assumed to be the same as the number of devices at the UEs node, i.e., $U$. Meanwhile, the number of data streams transmitted from the IAB donor and IAB-node is assumed to be $U$ as well. However, since each user is assumed to have one RF chain receiving one data stream, only analog beamforming is required. For the FR2 band, the Orthogonal Frequency Division Multiplexing (OFDM) system is adopted, where we assume i) the length of the data block is the same as the number of subcarriers, i.e., ${K}$; ii) the RF beamformers are frequency-flat and the same for all subcarriers. In contrast, the baseband (BB) beamformers are different for different subcarriers \cite{7448873}. The beamforming structure for this wideband FR2-IBFD-IAB network is shown in Fig.~\ref{sys}.

At the IAB donor, $U$ data streams are transmitted through $U$ RF chains and ${N_T}$ transmit antenna arrays. The total number of antenna arrays are equally divided into $U$ subarrays, each with one RF chain. Hence, the transmitted signal at the $k=1,2,\dotsc,{K}$th subcarrier from the IAB donor is given by
\begin{equation}
\mathbf{x}_\mr{D}[k] = \mathbf{F}_{\rf\mr{D}}\left(\underbrace{\mathbf{F}_{\bb\mr{D}}[k]\mathbf{s}_\mr{D}[k]}_{\widetilde{\mathbf{x}}_\mr{D}[k]}+\mathbf{e}_\mr{D}[k]\right),\label{xr}
\end{equation}
where $\mathbf{F}_{\rf\mr{D}}=\mr{blkdiag}\left[\mathbf{f}_{\rf\mr{D},1}, \mathbf{f}_{\rf\mr{D},2}, \ldots, \mathbf{f}_{\rf\mr{D},U}\right]\in\mathbb{C}^{N_{T} \times U}$ is the block diagonal RF precoder matrix with $\mathbf{f}_{\rf\mr{D},u}\in\mathbb{C}^{\frac{N_{T}}{U}\times1},\forall u\in\{1,2,\ldots,U\}$ representing the RF precoder vector of the $u$th subarray. $\mathbf{F}_{\bb\mr{D}}[k]\in \mathbb{C}^{U\times U}$ represents the BB precoder matrix. The transmit data vector $\mathbf{s}_\mr{D}[k] \in \mathbb{C}^{U \times 1}$ at the subcarrier $k$ has the covariance matrix of $\mathbb{E}\left\{\mathbf{s}_\mr{D}[k]\mathbf{s}^H_\mr{D}[k]\right\}=\frac{P_t}{KU}\mathbf{I}_{U}$, where $P_t$ is the average total transmit power across all subcarriers. By applying the transmit power constraint with equal power allocation, we get the constraint on the precoder as $\lVert\mathbf{F}_{\rf\mr{D}}\mathbf{F}_{\bb\mr{D}}[k]\rVert_F^2=U$ for all subcarriers. The vector $\mathbf{e}_\mr{D}[k]\in\mathbb{C}^{U\times1}\sim \mathcal{CN}\left(\mathbf{0},\rho\mr{diag} \left[\mr{Cov}\left[\widetilde{\mathbf{x}}_\mr{D}[k]\right]\right]\right)$ captures the transmitter HWI at the IAB donor with $\rho<<1$, where the transmitter HWI is uncorrelated with the transmit signal.

At the IBFD-IAB-node, separate antennas are configured for transmission and reception (i.e., there are ${n}_{{T}}$ transmit antenna arrays and $U$ RF chains for transmitting to the UEs node; and ${n}_{{R}}$ antenna arrays with $U$ RF chains for receiving data from the IAB donor). Similarly, the subarray structure divides those antenna arrays into $U$ equal panels, each with one RF chain. Without SIC, the decoded signal at the IAB-node for subcarrier $k$ is expressed in \eqref{yr}, shown on the top of next page,
\newcounter{mytempeqncnt}
\begin{figure*}[t!]
\normalsize
\setcounter{mytempeqncnt}{\value{equation}}
\setcounter{equation}{1}
\begin{align}
\mathbf{y}_\mr{N}[k]=\mathbf{W}_{\bb\mr{N}}^{H}[k]\left[\underbrace{\mathbf{W}_{\rf\mr{N}}^{H}\left(\mathbf{H}_\mr{ND}[k]\mathbf{x}_\mr{D}[k]+\mathbf{H}_\mr{SI}[k]\mathbf{x}_\mr{N}[k]+\mathbf{z}_\mr{N}[k]\right)}_{\widetilde{\mathbf{y}}_\mr{N}[k]}+\mathbf{g}_\mr{N}[k]\right]\label{yr}
\end{align}
\hrulefill
\setcounter{equation}{\value{mytempeqncnt}}
\end{figure*}
where $\mathbf{W}_{\rf\mr{N}}=\mr{blkdiag}\left[\mathbf{w}_{\rf\mr{N},1},\mathbf{w}_{\rf\mr{N},2}, \ldots, \mathbf{w}_{\rf\mr{N},U}\right]\in\mathbb{C}^{{n}_{R}\times U}$ represents the RF combiner matrix with $\mathbf{w}_{\rf\mr{N},u}\in\mathbb{C}^{\frac{n_{R}}{U}\times 1},\forall u\in\{1,2,\ldots,U\}$ denoting the RF combiner vector of subarray $u$.  $\mathbf{W}_{\bb\mr{N}}[k]\in\mathbb{C}^{U \times U}$ is the BB combiner matrix. $\mathbf{H}_\mr{ND}[k]\in\mathbb{C}^{{n}_{{R}} \times {N}_{{T}}}$ and $\mathbf{H}_\mr{SI}[k]\in\mathbb{C}^{{n}_{{R}} \times {n}_{{T}}}$ are the ideal backhaul channel matrix and SI channel matrix at the subcarrier $k$, respectively. $\mathbf{z}_\mr{N}[k] \in \mathbb{C}^{n_R \times 1}\sim \mathcal{CN}(\mathbf{0},\sigma^2_\mr{N}\mathbf{I}_{n_R})$ is the circularly symmetric Gaussian noise. The vector $\mathbf{g}_\mr{N}[k]\in\mathbb{C}^{U\times1}\sim \mathcal{CN}(\mathbf{0},\beta \mr{diag}[\mr{Cov}[\widetilde{\mathbf{y}}_\mr{N}[k]]])$ accounts for the receiver HWI at the IAB-node, which is uncorrelated with the received signal, with $\beta<<1$.

The vector $\mathbf{x}_\mr{N}[k]$ in \eqref{yr} denotes the signal transmitted from the IAB-node at the $k$th subcarrier, given as
\setcounter{equation}{2}
\begin{equation}
\mathbf{x}_\mr{N}[k] = \mathbf{F}_{\rf\mr{N}}\left(\underbrace{\mathbf{F}_{\bb\mr{N}}[k]\mathbf{s}_\mr{N}[k]}_{\widetilde{\mathbf{x}}_\mr{N}[k]}+\mathbf{e}_\mr{N}[k]\right),\label{xd}
\end{equation}
where $\mathbf{F}_{\rf\mr{N}}=\mr{blkdiag}\left[\mathbf{f}_{\rf\mr{N},1}, \mathbf{f}_{\rf\mr{N},2}, \ldots, \mathbf{f}_{\rf\mr{N},U}\right]\in\mathbb{C}^{n_{T} \times U}$ is the RF precoder matrix with $\mathbf{f}_{\rf\mr{N},u}\in\mathbb{C}^{\frac{n_{T}}{U}\times 1},\forall u\in\{1,2,\ldots,U\}$ denoting the RF precoder vector of the $u$th subarray. $\mathbf{F}_{\bb\mr{N}}[k]=\left[\mathbf{f}_{\bb\mr{N},1}[k],\mathbf{f}_{\bb\mr{N},2}[k],\ldots,\mathbf{f}_{\bb\mr{N},U}[k]\right]\in \mathbb{C}^{U \times U}$ represents the BB precoder matrix with $\mathbf{f}_{\bb\mr{N},u}[k]\in\mathbb{C}^{U\times 1},\forall u\in\{1,2,\ldots,U\}$. $\mathbf{s}_\mr{N}[k] \in \mathbb{C}^{U\times 1}$ is the transmit data vector with covariance matrix of $\mathbb{E}\left\{\mathbf{s}_\mr{N}[k]\mathbf{s}^H_\mr{N}[k]\right\}=\frac{P_t}{KU}\mathbf{I}_U$ and is uncorrelated with $\mathbf{s}_\mr{D}[k]$. The vector $\mathbf{e}_\mr{N}[k]\in\mathbb{C}^{U\times1}\sim \mathcal{CN}\left(\mathbf{0},\rho\mr{diag} \left[\mr{Cov}\left[\widetilde{\mathbf{x}}_\mr{N}[k]\right]\right]\right)$ denotes the transmitter HWI at the IAB node, which is uncorrelated with the transmit signal. In addition, for all subcarriers, the precoder per subarray has to satisfy the constraint of $\lVert \mathbf{F}_{\rf\mr{N}}\mathbf{f}_{\bb\mr{N},u}[k] \rVert_F^2=1,\forall u\in\{1,2,\ldots,U\}$ for sending data stream to the $u$th UE.

At the UEs node, there are $U$ devices, each is equipped with $N_R$ receive antennas and a single RF chain. Thus, the received signal at \textit{all UEs} can be jointly written as
\begin{equation}
\mathbf{y}_\mr{E}[k]=\underbrace{\mathbf{W}_{\rf\mr{E}}^H\left(\mathbf{H}_\mr{EN}[k]\mathbf{x}_\mr{N}[k]+\mathbf{z}_\mr{E}[k]\right)}_{\widetilde{\mathbf{y}}_\mr{E}[k]}+\mathbf{g}_\mr{E}[k],\label{yd}
\end{equation}
where $\mathbf{y}_\mr{E}[k]=\left[y_{\mr{E},1}[k],y_{\mr{E},2}[k],\ldots,y_{\mr{E},U}[k]\right]^T\in\mathbb{C}^{U\times1}$ with $y_{\mr{E},u}[k],\forall u\in\{1,2,\ldots,U\}$ denoting the decoded signal at the $u$th UE. $\mathbf{W}_{\rf\mr{E}}=\mr{blkdiag}\left[\mathbf{w}_{\rf\mr{E},1},\mathbf{w}_{\rf\mr{E},2}, \ldots,\mathbf{w}_{\rf\mr{E},U}\right]\in\mathbb{C}^{UN_R\times U}$ is the RF combiner matrix with $\mathbf{w}_{\rf\mr{E},u}\in\mathbb{C}^{{N_R}\times1},\forall u\in\{1,2,\ldots,U\}$ being the RF combiner vector of the $u$th UE. $\mathbf{H}_\mr{EN}[k]=\left[\mathbf{H}_{\mr{EN},1}^T[k],\mathbf{H}_{\mr{EN},2}^T[k],\ldots,\mathbf{H}_{\mr{EN},U}^T[k]\right]^T\in\mathbb{C}^{UN_R\times{n_T}}$ is the ideal access link channel matrix, where $\mathbf{H}_{\mr{EN},u}[k]\in\mathbb{C}^{{N}_{R}\times{n}_{T}}$ represents the access link channel matrix from the IAB-node to the $u$th UE. $\mathbf{z}_\mr{E}[k]=\left[\mathbf{z}_{\mr{E},1}^T[k],\mathbf{z}_{\mr{E},2}^T[k],\ldots,\mathbf{z}_{\mr{E},U}^T[k]\right]^T \in \mathbb{C}^{{UN}_{R} \times 1} \sim \mathcal{CN}(\mathbf{0},\sigma^2_\mr{E}\mathbf{I}_{UN_{R}})$ is the Gaussian noise vector with $\mathbf{z}_{\mr{E},u}[k]\in\mathbb{C}^{n_R \times 1},\forall u\in\{1,2,\ldots,U\}$ being the Gaussian noise vector at the $u$th UE. The receiver HWI vector $\mathbf{g}_\mr{E}[k]=[g_{\mr{E},1}[k],g_{\mr{E},2}[k],\ldots,$ $g_{\mr{E},U}[k]]^T\in\mathbb{C}^{U\times1}\sim \mathcal{CN}(\mathbf{0},\beta \mr{diag}[\mr{Cov}[\widetilde{\mathbf{y}}_\mr{E}[k]]])$ with $g_{\mr{E},u}[k],\forall u\in\{1,2,\ldots,
U\}$ denoting the receiver HWI at the $u$th UE, which is uncorrelated with the received signal.

\subsection{General Channel}
For the wideband FR2 communications with the OFDM system, a cyclic prefix of length $D$ is added to each OFDM symbol, which is equal to the number of delay taps for the wideband channel. Due to the scattering effect, the FR2 signals are likely to arrive in ${N_C}$ clusters, with ${N_L}$ paths reflected by different obstacles in each cluster. A raised-cosine pulse shaping filter $p(dT_s-\tau_{c,l})$, for $d=0,1,\ldots,D-1$, with $T_s$-spaced signaling is utilized, where the delay $\tau_{c,l}$ is defined for the $l$th path in the $c$th cluster \cite{7094443}. Assuming uniform planar arrays (UPAs) with half-wavelength spaced elements, the transmit and receive steering vectors can be written as $\mathbf{a}_{\mr{t}}(\theta_{c,l}^t,\phi_{c,l}^t)$ and $\mathbf{a}_{\mr{r}}(\theta_{c,l}^r,\phi_{c,l}^r)$, respectively, where the azimuth $\theta_{c,l}^r/\theta_{c,l}^t$ and elevation $\phi_{c,l}^r/\phi_{c,l}^t$ angles correspond to the angles of arrival/departure (AoAs/AoDs) for each path in their clusters. Hence, at subcarrier $k$, a typical FR2 channel model between two nodes can be expressed as
\begin{equation}
    \mathbf{H}[k]=\mathbf{A}_r\mathbf{\Pi}[k]\mathbf{A}_t^H,\label{channel}
\end{equation}
where
\begin{equation}
    \mathbf{A}_r=[\mathbf{a}_{\mr{r}}(\theta_{1,1}^r,\phi_{1,1}^r),\ldots,\mathbf{a}_{\mr{r}}(\theta_{c,l}^r,\phi_{c,l}^r),\ldots,\mathbf{a}_{\mr{r}}(\theta_{{N}_{C},{N}_L}^r,\phi_{{N}_{C},{N}_L}^r)],
\end{equation}
\begin{equation}
    \mathbf{A}_t=[\mathbf{a}_{\mr{t}}(\theta_{1,1}^t,\phi_{1,1}^t),\ldots,\mathbf{a}_{\mr{t}}(\theta_{c,l}^t,\phi_{c,l}^t),\ldots,\mathbf{a}_{\mr{t}}(\theta_{{N}_{C},{N}_L}^t,\phi_{{N}_{C},{N}_L}^t)],
\end{equation}
\begin{equation}
\mathbf{\Pi}[k]=\sqrt{\tfrac{{N}_{r} {N}_{t} }{{N}_{C} {N}_{L}\bar{PL}}}\begin{bmatrix}\begin{smallmatrix}
  \alpha_{1,1}\chi_{1,1}[k] &&\hdots&& 0\\\vdots&&\ddots&&\vdots\\0&\hdots&\alpha_{c,l}\chi_{c,l}[k]&\hdots&0\\\vdots&&\ddots&&\vdots\\0 &&\hdots&& \alpha_{{N}_{C},{N}_L}\chi_{{N}_{C},{N}_L}[k]
 \end{smallmatrix}
  \end{bmatrix},
\end{equation}
and $\chi_{c,l}[k]=\sum_{d=0}^{D-1}p(dT_s-\tau_{c,l})e^{(-j\frac{2{\pi}kd}{K})}$. ${N}_{t}$ and ${N}_{r}$ denote the number of transmit and receive antennas, respectively. $\alpha_{c,l}$ is the complex gain. $\bar{PL}$ denotes the average path loss due to the high attenuation of FR2 band channel. The close-in path loss model is adopted rather than the free space path loss \cite{7414036}, given as
\begin{align}
\bar{PL}=\left(\frac{4\pi r_0}{\lambda}\right)^2\left(\frac{r}{r_0}\right)^\mu,\label{pathloss}
\end{align}
where $r_0$, $r$, $\lambda$, and $\mu$ represent the reference distance, distance between transceiver, wavelength, and path loss exponent, respectively. Moreover, since for arbitrary transmission networks, the line-of-sight (LOS) component has a high probability of being blocked by obstacles. Therefore, an non-line-of-sight (NLOS) path loss exponent is preferred. Furthermore, the steering vector is defined as
\begin{equation}
\mathbf{a}(\theta,\phi)=\tfrac{1}{\sqrt{N}}\big[1,a_1(\theta,\phi),\dotsc,a_{N-1}(\theta,\phi)]^{T},
\label{steering}
\end{equation}
where $a_n(\theta,\phi)=e^{j\frac{2\pi}{\lambda}\mathbf{r}_n^{T}\mathbf{u}(\theta,\phi)}$; $N$ is the number of antenna arrays in the UPA;
$\mathbf{r}_n=[x_n,y_n,z_n]^T$ is the coordinate of the $n$th antenna element; $\mathbf{u}(\theta,\phi)=[\cos\theta\cos\phi,\sin\theta\cos\phi,\sin\phi]^{T}$ is the unit-norm direction vector. In this work, the arrays are placed in the XY-plane, and the elevation angles are measured from the XY-plane. Besides, the $z$-axis indicates the array height measured from the UPA plane, which is assumed to be negligible, i.e., $z_n\approx0$.

\subsection{Self-Interference Channel}
The most important issue in the IBFD transmission is the introduction of the SI on the IAB-node. Due to the proximity of the transceiver on the IAB-node, the attenuation of the SI channel is significantly less than that of the typical communication channels, which contributes high power SI to the backhaul link and degrades its SE. In order to reduce the effect of the SI, a staged SIC scheme will be introduced in later sections.

Since the distinct SI channel model for the FR2 band is still unknown, most of the works have considered the hypothetical SI channel for narrowband communications \cite{8246856,1234}. Fortunately, a hypothetical model is proposed for the wideband SI channel in \cite{9149419}. According to \cite{1234,9149419}, after some minor modifications, we model the hypothesis wideband SI channel as follows. Unlike the general channel in the previous subsection, the SI channel is likely to be modeled as a Rician-alike channel with Rician factor $\kappa$. The LOS part, $\mathbf{H}_\mr{SI,L}$, is adopted to a near-field model with spherical waveform and is assumed to be frequency flat. The frequency response of the LOS component is given as
\begin{align}
\mathbf{H}_\mr{SI,L}=\left[\mathbf{a}_r(\theta^r,\phi^r)\mathbf{a}^H_t(\theta^t,\phi^t)\right]\odot\mathbf{R},
\end{align}
where only one AoA/AoD is assumed for the LOS link. The entries of $\mathbf{R}$ is $[\mathbf{R}]_{p,q}=\frac{\gamma}{{r}_{pq}}e^{-j 2 \pi \frac{{r}_{pq}}{\lambda}}$ with ${r}_{pq}$ denoting the distance between the $p$th element of the receive antenna and the $q$th element of the transmit antenna at the IAB-node. $\gamma=\sqrt{n_Rn_T}$ is the normalization factor ensuring that the norm of $\mathbf{H}_\mr{SI,L}$ remains the same before and after multiplying with the steering vectors.

The NLOS part, $\mathbf{H}_\mr{SI,N}$, is expressed similar to the general channel model in \eqref{channel}, but with a few clusters and rays. Consequently, the entire SI channel for subcarrier $k$ can be expressed as
\begin{equation}
\mathbf{H}_\mr{SI}[k]=\sqrt{\frac{\kappa}{\kappa+1}}\mathbf{H}_\mr{SI,L}+ \sqrt{\frac{1}{\kappa+1}}\mathbf{H}_\mr{SI,N}[k].
\end{equation}

\section{Analog Self-Interference Cancellation}\label{asic}
In this section, the working principle and limiting factor of the conventional A-SIC idea are presented first. Then, the OD-based canceler is described, followed by the implementation details of such canceler design for the FR2-IBFD-IAB networks. In this work, we assume the antenna isolation has already been deployed before A-SIC.

\subsection{Working Principle and Limitations}
A-SIC is essential to avoid receiver saturation. Otherwise, the signal-of-interest cannot be quantized precisely \cite{7922557,8171089}. Active A-SIC is based on a subtraction idea, i.e., a replica of the received SI signal generated by the analog canceler is inserted into the receiver chain to subtract the received SI. The canceler is made up of limited number of tunable delay lines to capture the multi-path nature of the SI channel, where passive components are utilized to construct tunable delay lines to minimize the non-linearity effects. With multi-tap RF canceler, one can cancel the SI from reflection paths in addition to the direct path. By considering the hardware insertion losses, the frequency response of a single multi-tap RF canceler can be given as
\begin{equation}
\label{eq_canc_fre}
h_\mr{can}[\omega]=\hat{\alpha}\sum_{m=1}^{M}\alpha _m\beta _m\left(w_{I,m}+jw_{Q,m}\right)e^{-j \omega \tau_m},
\end{equation}
where $\hat{\alpha}$ is the attenuation introduced by coupling the RF signal into the canceler; $\alpha _m$ is the propagation loss of each delay line; $\beta _m$ denotes the tap coupling factor \cite[(4)]{8796422}; $w_{I,m}$ and $w_{Q,m}$ are tunable weights; and $\tau_m$ is the delay. The optimal weights are tuned to minimize the difference between the frequency components of the canceler and the SI channel within the band of interest (BoI) (for details, see \cite{8171089}). Equation \eqref{eq_canc_fre} suggests that the number of taps $M$ decides the available degrees of freedom for this optimization. The key factor for efficient wideband A-SIC is the realization of a sufficient number of taps (i.e., delay lines) \cite{luo}. For wider operational bandwidth, more frequency components need to be optimized, and more degrees of freedom, i.e., taps, are required.

\begin{figure*}[t!]
\centering
\includegraphics[width=\textwidth]{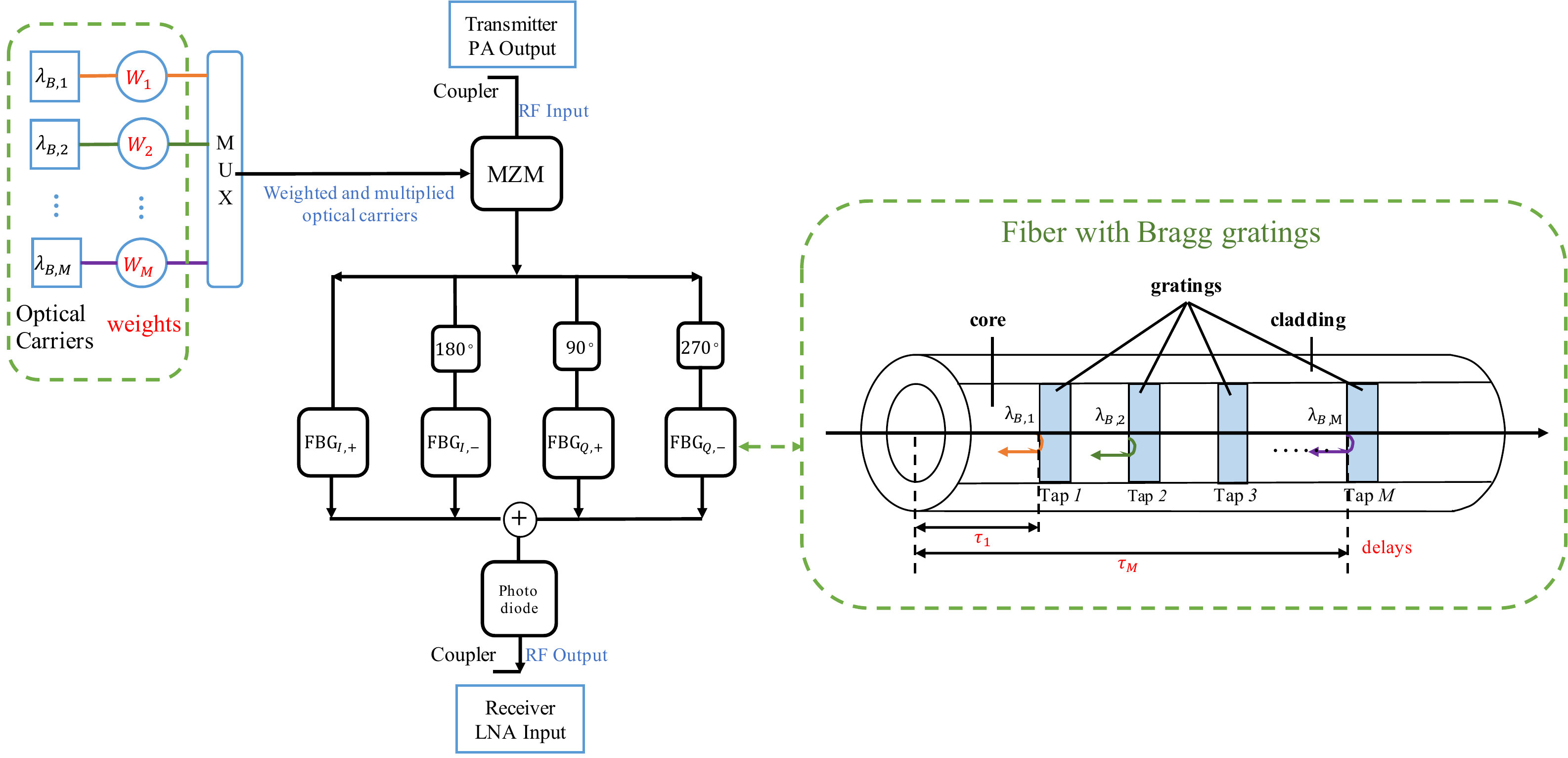}
\caption{Illustration of the OD-based analog canceler.}
\label{principle_FBG}
\end{figure*}

\subsection{OD-Based A-SIC}
For the conventional canceler, the insertion losses increase with an increasing number of taps (i.e., $\alpha_m$ and $\beta_m$ are small for large $m$ in \eqref{eq_canc_fre}), which results in a large difference between the signal power at the first and the later taps. Therefore, the signals coupled into the later taps cannot replicate the desired signal level and degrade the cancellation performance. Conventionally, electrical attenuators and micro-strips or cables can be used for constructing the tunable delay lines. However, it is demonstrated that these electrical components have significant propagation loss and coupling loss that limit the number of effective taps \cite{8796422}, thus limiting the operational bandwidth and cancellation performance. Therefore, to overcome these drawbacks, an OD-based analog canceler has recently been investigated in \cite{8796422}, whose structure is illustrated in Fig.~\ref{principle_FBG}.

Regarding the OD-based canceler mechanism, the RF reference signal is first converted to the optical domain by modulating onto optical carriers through the Mach–Zehnder modulator (MZM). These optical carriers are generated by tunable lasers according to the grating wavelengths, and the power of these carriers is adjusted by variable optical attenuators (VOAs). Then, $M$ optical carriers are combined by a multiplexer (MUX) for propagating into a single fiber according to the obtained weights. The reference signal modulated on the optical carrier at wavelength $\lambda_{B,m}$ will be reflected at the $m$th grating while propagating through the fiber-Bragg-grating (FBG). This reflection happens at different gratings causes different time delays to the coupled reference signal. Next, the reflected signals are detected by photo-diodes to remove the optical carriers. Finally, the canceler yields an accumulation of multiple weighted and delayed versions of the input reference signal as the canceler output \cite{8796422}. Since the weights are achieved by attenuators, which can only be real and non-negative; however, the SI channel is complex. Thus, four FBGs are needed to realize the complex response of the canceler.

The OD-based canceler can also be described by \eqref{eq_canc_fre}. Compared with conventional canceler, OD-based canceler has smaller insertion losses, i.e., $\alpha_m$ and $\beta_m$ are almost constant with increasing $m$. Theoretically, almost constant insertion losses in the OD-based canceler allow hundreds of effective taps to be implemented to enlarge the operational bandwidth.

\subsection{Proposed OD-Based A-SIC}
In order to realize the OD-based canceler design in the MIMO system, ${n_R}\times {n_T}$ cancelers are traditionally required to match the ${n_R}\times {n_T}$ SI channel matrix, where each canceler is constructed and tuned as described above. However, such a canceler deployment will be extremely costly for the FR2 communications, especially for the OD-based canceler. In order to reduce the cost, we tap off the SI signal from the RF chains before the RF precoder at the IAB-node transmitter and insert the outputs of these analog cancelers back to the RF chains at the IAB-node receiver after the RF combiner (see Fig~\ref{sys}) \cite{roberts2020equipping}. With this architecture, the required number of analog cancelers can be reduced from ${n_R}\times {n_T}$ to $U\times U$, which is of great benefit to the cost and practical implementation. Since a single canceler can be tuned by adjusting the weights to imitate the estimated RF SI channel $\hat{h}_{\mr{SI},pq}[\omega]$ between the $p$th transmitter's RF chain to the $q$th receiver's RF chain, where $p,q\in\{1,2,\ldots,U\}$, the following optimization problem will need to be run for each canceler established between the $pq$th RF chain pair over the BoI, which is cast as
\begin{equation}
    \begin{aligned}
&\arg\min_{\left\{w_{I,m}^{pq},\; w_{Q,m}^{pq}\right\}_{m=1}^M} \,\, \sum_{p=1}^U \sum_{q=1}^U \left\{\left \| \hat{h}_{\mr{SI},pq}[\omega]-h_{\mr{can},pq}[\omega] \right \|^2\right\}_{\omega=\omega_0}^{\omega_1} \\
&\text{ s.t.}\quad
-1 \leq w_{I,m}^{pq} \leq 1,\, -1 \leq w_{Q,m}^{pq} \leq 1,
    \end{aligned}
\end{equation}
where $[{\omega_0},{\omega_1}]$ spans the BoI, i.e., [27.8 GHz, 28.2 GHz] in this work. The operator $\{\left \| \cdot \right \|^2\}_{\omega=\omega_0}^{\omega_1}$ means the sum of the squared error across all frequency components within $[{\omega_0},{\omega_1}]$ since the sampled version of the BoI is considered \cite{luo}. $h_{\mr{can},pq}[\omega]$ is the canceler response for mitigating the SI between the $pq$th transceiver RF chain pair, which is represents by \eqref{eq_canc_fre}. The constraints come from passive VOAs. With the channel state information (CSI) of the estimated RF SI channel and the frequency response of the canceler without VOA effects being known as a prior, the optimal weights can be obtained by the least-squares (LS) method.

Since the A-SIC performance mainly depends on the frequency selectivity of the SI channel and the number of taps in the canceler, and due to the fact that the RF beamformers do not affect the frequency selectivity of the SI channel, we assume the amount of cancellation for the RF SI channel to be the same as that for the SI channel. Thus, we obtain the A-SIC performance through simulating with the SI channel instead of the RF SI channel and reflect the A-SIC effect by simply scaling the SI signal with a power attenuation factor.

In this work, we assume antenna isolation also attenuates the SI signal in a frequency-flat manner \cite{9124849}. Thus, after A-SIC, the term $\mathbf{H}_\mr{SI}[k]\mathbf{x}_\mr{N}[k]$ in \eqref{yr} is scaled by $\sqrt{\eta}$ with the scalar $\eta$ being the amount of SI signal strength attenuated by both the antenna isolation and A-SIC.

\section{RF Codebook Design and RF Effective Channel Estimation}\label{RFest}
In practice, the RF precoders/combiners are usually implemented using finite resolution PSs, i.e., they are selected from the pre-defined RF codebooks. Besides, the estimation of the large and sparse mmWave channel is difficult in reality. Motivated by these, in this section, a modified LBG algorithm will be introduced for designing the RF codebook, followed by the estimation of the RF effective channels after A-SIC.

\subsection{Modified MSE-Based LBG Algorithm for RF Codebook Design}
The LBG algorithm is a popular vector quantization scheme and is treated as an extension of the Lloyd-Max scalar quantization algorithm \cite{1094577}. Conventionally, for a matrix quantization, the existing codebooks work by vector-wise comparison can lead to a low-rank behavior on the quantized matrix\footnote[2]{Suppose each subarray has multiple RF chains. With a vector-wise codebook, likely, the columns for the RF beamforming matrix of a certain subarray may be assigned to the same vector codeword, which can result in a low-rank matrix and the loss of degrees of freedom.}. Therefore, to avoid that, we modify the LBG algorithm to yield the $B$ bits codebook with matrix codewords directly, whose steps are described as follows.
\begin{itemize}
    \item \textit{Step 1} (Initialization):\\
    Given the training set $\mathcal{F}=\big\{\mathbf{F}_{\rf,t}|t=1,2,\ldots,T, \left|\mathbf{F}_{\rf,t}\right|_{pq}=1$ if $\left[\mathbf{F}_{\rf,t}\right]_{p,q}\neq0\big\}$ with $T$ entries, whose each entry is a block diagonal matrix with each block denoted by the angle of complex Gaussian random numbers with zero mean and unit variance\footnote[3]{Optimally, the training set should have consisted of the optimal RF precoders/combiners, which are derived by the angle of the dominant eigenvector(s) corresponding to the eigenvalue decomposition (EVD) of the channel correlation matrix (i.e., the sample covariance matrix) \cite{7880698}. However, as aforementioned, the mmWave channel is hard to be estimated. Therefore, by exploring the distribution of the RF precoders/combiners, i.e., the values in the RF precoder/combiner matrix are isotropically (uniformly) distributed \cite[Lemma 1, 2]{8464682,1237152}, we construct the entries of the training set by the angle of $\mathcal{CN}(0,1)$ random numbers}. The codebook $\mathcal{C}$ is initialized with an entry $\mathbf{C}_1(0)$, obtained by the angle of the mean value of the training set as
    \begin{equation}
        \mathbf{C}_1(0)=e^{j \arg\left(\frac{\sum_{t=1}^T\mathbf{F}_{\rf,t}}{T}\right)}.
    \end{equation}
    \item \textit{Step 2} (Splitting):\\
    This step splits each entry of the $b$ bits codebook $\mathcal{C}$ into two new ones to initialize the $b+1$ bits codebook, where $b=0,1,\ldots,B-1$. To achieve that, we perturb each entry $\mathbf{C}_i(b)$ as,
    \begin{align}
        &\mathbf{C}^{(0)}_{i+2^b}(b+1)=e^{j\arg\left(\sqrt{1-\epsilon^2}\mathbf{C}_i(b)+\epsilon\mathbf{P}_i(b)\right)},\notag\\&\mathbf{C}^{(0)}_{i}(b+1)=e^{j\arg\left(\sqrt{1-\epsilon^2}\mathbf{C}_i(b)-\epsilon\mathbf{P}_i(b)\right)},
    \end{align}
    where $i=1,2,\ldots,2^b$, $\epsilon$ is a small positive value (e.g., $10^{-3}$), $\mathbf{P}_i(b)$ is a block diagonal matrix, whose each block is drawn from the angle of $\mathcal{CN}(0,1)$ random numbers.
    \item \textit{Step 3} (Cluster Assignment):\\
    In this step, using the nearest neighbor routine based on MSE, the training set is divided into $2^{b+1}$ (i.e., $|\mathcal{C}|$) clusters, the centroid of cluster $j$ is given by $\mathbf{C}_j^{(v)}(b+1)$, where $v=0,1,\ldots,V-1$ with $V$ being the maximum number of iterations of Step 5. E.g., $\mathbf{F}_{\rf,t}$ is in the cluster 1 if $d\left(\mathbf{F}_{\rf,t},\mathbf{C}_1^{(v)}(b+1)\right)\leq d\left(\mathbf{F}_{\rf,t},\mathbf{C}_j^{(v)}(b+1)\right)$, $\forall j=1,2,\ldots,|\mathcal{C}|$, where $d\left(\mathbf{X},\mathbf{Y}\right)=\frac{1}{PQ}\sum_{p=1}^P\sum_{q=1}^Q\left([\mathbf{X}]_{p,q}-[\mathbf{Y}]_{p,q}\right)^2$, and $P$, $Q$ denote the number of rows and columns of the matrix, respectively.
    \item \textit{Step 4} (Centroid Update):\\
    Each entry of the codebook is updated with the centroid of the corresponding cluster. The centroid is computed via the solution of the following optimization problem, that is
    \begin{equation}
        \mathbf{{C}}_j^{(v)}(b+1)=\arg\underset{\mathbf{C}_j^{(v)}(b+1)}\min{\underset{\mathbf{F}_{\rf,t}\in j}\sum d\left(\mathbf{F}_{\rf,t},\mathbf{C}_j^{(v)}(b+1)\right)}.
    \end{equation}
    Thus, the new centroid $\mathbf{{C}}_j^{(v)}(b+1)$ is given by the angle of the mean value of all $\mathbf{F}_{\rf,t}$ in the $j$th cluster.
    \item \textit{Step 5} (Inner loop):\\
    Go to step 3 until the maximum number of iterations $V$ is reached (e.g., $V=50$).
    \item \textit{Step 6} (Outer loop):\\
     Go to step 2 until the length of the codebook $b+1$ is equal to the desired  codebook length $B$.
\end{itemize}

\subsection{RF Effective Channel Estimation}\label{RFCO}
Given the RF codebooks, we can estimate the RF effective channels for designing the BB beamformers. Note that the RF effective channels estimated in this section are those after A-SIC by assuming BB beamformers to be identity matrices \cite{9431171}. 

There are two phases in the RF effective channel estimation: i) RF precoder-combiner pair selection; ii) RF effective channel estimation. The RF beamformers are designed to maximize the desired signal in their corresponding links. We treat the whole OFDM symbols as pilots and assume only the IAB donor or the IAB node can transmit data in a time slot. Moreover, the identity BB beamformer matrices are omitted here.

\subsubsection{Phase 1 (RF precoder-combiner pair selection)} 
The received backhaul link signal of the $k$th pilot subcarrier at the IAB-node, which uses the $p$th codeword of the codebook $\mathcal{F}_\mr{D}$ as the RF precoder and the $q$th codeword of the codebook $\mathcal{W}_\mr{N}$ as the RF combiner, is given by
\begin{align}
\mathbf{Y}_\mr{N}[k](p,q)&= \mathbf{W}_{\rf\mr{N},q}^H[\mathbf{H}_\mr{ND}[k]\mathbf{F}_{\rf\mr{D},p}\left(\mathbf{S}_\mr{D}[k]+\mathbf{E}_\mr{D}[k]\right)\notag\\&+\mathbf{Z}_\mr{N}[k]]+\mathbf{G}_\mr{N}[k],
\end{align}
where $\mathbf{S}_\mr{D}[k]\in\mathbb{C}^{U\times U}$ is the matrix of orthogonal pilot signal with $\mathbf{S}_\mr{D}[k]\mathbf{S}_\mr{D}^H[k]=\frac{P_t}{KU}\mathbf{I}_{U}$. $\mathbf{E}_\mr{D}[k]\in\mathbb{C}^{U\times U}$, $\mathbf{G}_\mr{D}[k]\in\mathbb{C}^{U\times U}$, and $\mathbf{Z}_\mr{N}[k]\in\mathbb{C}^{U\times U}$ are the noise matrices caused by the transmitter HWI, receiver HWI, and Gaussian noise, respectively, following the same statistics in  \eqref{xr} and \eqref{yr}.

Similarly, the jointly received access link signal of the $k$th pilot subcarrier \textit{across all UEs}, which uses the $p$th codeword of the codebook $\mathcal{F}_\mr{N}$ as the RF precoder and the $q$th codeword of the codebook $\mathcal{W}_\mr{E}$ as the RF combiner, is cast as
\begin{align}
\mathbf{Y}_\mr{E}[k](p,q)&= \mathbf{W}_{\rf\mr{E},q}^H[\mathbf{H}_\mr{EN}[k]\mathbf{F}_{\rf\mr{N},p}\left(\mathbf{S}_\mr{N}[k]+\mathbf{E}_\mr{N}[k]\right)\notag\\&+\mathbf{Z}_\mr{E}[k]]+\mathbf{G}_\mr{E}[k],
\end{align}
where the matrix of orthogonal pilot signal $\mathbf{S}_\mr{N}[k]\in\mathbb{C}^{U\times U}$ has $\mathbf{S}_\mr{N}[k]\mathbf{S}_\mr{N}^H[k]=\frac{P_t}{KU}\mathbf{I}_{U}$. $\mathbf{E}_\mr{N}[k]\in\mathbb{C}^{U\times U}$, $\mathbf{G}_\mr{E}[k]\in\mathbb{C}^{U\times U}$, and $\mathbf{Z}_\mr{E}[k]\in\mathbb{C}^{U\times U}$ are the transmitter HWI, receiver HWI, and Gaussian noise matrix, respectively, with the same statistics in \eqref{xd} and \eqref{yd}.

According to the beam management \cite{7947209}, each time, a codeword is chosen from their corresponding codebook and the RF precoder and combiner pairs that can maximize the received power among all pilot subcarriers are selected, given as
\begin{subequations}
\begin{alignat}{2}
& \left\{\mathbf{F}_{\rf\mr{D}},\mathbf{W}_{\rf\mr{N}}\right\}=\arg\underset{p,q}\max{\sum_{k=1}^K\left\|\mathbf{Y}_\mr{N}[k](p,q)\right\|_F^2} & &\\
& \text{subject to} \quad \mathbf{F}_{\rf\mr{D},p} \in \mathcal{F}_\mr{D},\quad\mathbf{W}_{\rf\mr{N},q} \in \mathcal{W}_\mr{N}.
\end{alignat}
\end{subequations}
\begin{subequations}
\begin{alignat}{2}
& \left\{\mathbf{F}_{\rf\mr{N}},\mathbf{W}_{\rf\mr{E}}\right\}=\arg\underset{p,q}\max{\sum_{k=1}^K\left\|\mathbf{Y}_\mr{E}[k](p,q)\right\|_F^2} & &\\
& \text{subject to} \quad \mathbf{F}_{\rf\mr{N},p} \in \mathcal{F}_\mr{N},\quad\mathbf{W}_{\rf\mr{E},q} \in \mathcal{W}_\mr{E}.
\end{alignat}
\end{subequations}
In this work, the RF beamformers for all nodes can be selected from the same isotropic RF codebook derived from the last subsection.
\subsubsection{Phase 2 (RF effective channel estimation)} Given the RF precoder/combiner, one can estimate the RF effective channel with the help of pilot signal by standard estimation methods, such as, the LS. 

Consequently, after estimation, we can write the ideal RF effective channel matrix as the sum of the estimated RF effective channel matrix $\hat{\mathbf{H}}^{eff}_{(\cdot)}[k]$ and the estimation error matrix $\mathbf{\Delta}_{(\cdot)}[k]$, given as 
\begin{equation}
    \mathbf{W}_{\rf\mr{N}}^H\mathbf{H}_\mr{ND}[k]\mathbf{F}_{\rf\mr{D}}=\hat{\mathbf{H}}^{eff}_\mr{ND}[k]+\mathbf{\Delta}_\mr{ND}[k],\label{RFB}
\end{equation}
\begin{equation}
    \sqrt{\eta}\mathbf{W}_{\rf\mr{N}}^H\mathbf{H}_\mr{SI}[k]\mathbf{F}_{\rf\mr{N}}=\hat{\mathbf{H}}^{eff}_\mr{SI}[k]+\mathbf{\Delta}_\mr{SI}[k],\label{RFSI}
\end{equation}
\begin{equation}
    \mathbf{W}_{\rf\mr{E}}^H\mathbf{H}_\mr{EN}[k]\mathbf{F}_{\rf\mr{N}}=\hat{\mathbf{H}}^{eff}_\mr{EN}[k]+\mathbf{\Delta}_\mr{EN}[k],\label{RFA}
\end{equation}
where we assume the channel estimation errors $\mathbf{\Delta}_\mr{ND}[k]$, $\mathbf{\Delta}_\mr{SI}[k]$, and $\mathbf{\Delta}_\mr{EN}[k]$ have the covariance matrices of $\mr{Cov}\left[\mathbf{\Delta}_\mr{ND}[k]\right]=\sigma_{e,\mr{ND}}^2\mathbf{I}_{M}$, $\mr{Cov}\left[\mathbf{\Delta}_\mr{SI}[k]\right]=\sigma_{e,\mr{SI}}^2\mathbf{I}_{M}$, and $\mr{Cov}\left[\mathbf{\Delta}_\mr{EN}[k]\right]=\sigma_{e,\mr{EN}}^2\mathbf{I}_{M}$ \cite{4686268,911302}.

\section{Digital Self-Interference Cancellation}\label{digital}
After A-SIC, the RSI left by previous stages will be processed in the digital domain of the IAB-node receiver. In practice, since the IAB-node knows its transmitted codeword $\mathbf{s}_\mr{N}[k]$ and we can know the estimated RF effective SI channel $\hat{\mathbf{H}}^{eff}_\mr{SI}[k]$ by the process in Section~\ref{RFest}. Then, with the help of successive interference cancellation, we can cancel out $\hat{\mathbf{H}}^{eff}_\mr{SI}[k]\mathbf{F}_\mr{\bb\mr{N}}[k]\mathbf{s}_\mr{N}[k]$.

Consequently, after subtraction, the decoded signal at the IAB-node in \eqref{yr} can be reconstructed as
\begin{align}
\hat{\mathbf{y}}_\mr{N}[k]=\mathbf{W}_{\bb\mr{N}}^H[k]\left({\widetilde{\hat{\mathbf{y}}}_\mr{N}[k]}+\mathbf{g}_\mr{N}[k]\right),\label{RFres}
\end{align}
where $\mathbf{W}_{\bb{\mr{N}}}[k]$ is designed to act as the minimum mean-squared error (MMSE) BB combiner, which will be described in the next section. $\widetilde{\hat{\mathbf{y}}}_\mr{N}[k]=\mathbf{W}_{\rf\mr{N}}^H\left(\mathbf{H}_\mr{ND}[k]\mathbf{x}_\mr{D}[k]+ \sqrt{\eta}\mathbf{H}_\mr{SI}[k]\mathbf{F}_{\rf\mr{N}}\mathbf{e}_\mr{N}[k]+\mathbf{z}_\mr{N}[k]\right)+\mathbf{\Delta}_\mr{SI}[k]\mathbf{F}_{\bb\mr{N}}[k]\mathbf{s}_\mr{N}[k]$

\section{Spectral Efficiency and Baseband Beamforming Design}\label{Sec4}
\subsection{Spectral Efficiency}
Define $\zeta=\frac{P_t}{KU}$ and substitute \eqref{xr}, \eqref{RFB}, and \eqref{RFSI} into \eqref{RFres}, the SE of the backhaul link is expressed according to \eqref{RFres}, given as
\begin{align}
\mathcal{R}_b&=\frac{1}{K}\sum_{k=1}^K\log_2\mr{det}\Big\{\mathbf{I}_{U}+{\mathbf{W}_{\bb{\mr{N}}}^H[k]\mathbf{\Phi}_{b}[k]\mathbf{W}_{\bb{\mr{N}}}[k]}\notag\\&\times\left(\mathbf{W}_{\bb{\mr{N}}}^H[k]\mathbf{\Omega}_{b}[k]\mathbf{W}_{\bb{\mr{N}}}[k]\right)^{-1}\Big\}, \label{backhaullink}
\end{align}
where $\mathbf{\Phi}_{b}[k]$ is the covariance matrix for the known part of the desired signal. $\mathbf{\Omega}_{b}[k]$ represents the covariance matrix consisting of the noise given by the channel estimation error, the transceiver HWI, and the Gaussian noise.
\begin{align}
    \mathbf{\Phi}_{b}[k]=\zeta\hat{\mathbf{H}}_\mr{ND}^{eff}[k]\mathbf{F}_\mr{\bb\mr{D}}[k]\mathbf{F}_\mr{\bb\mr{D}}^H[k]\left(\hat{\mathbf{H}}_\mr{ND}^{eff}[k]\right)^H.
\end{align}
\begin{align}
    \mathbf{\Omega}_{b}[k]&=\mathbf{\Omega}_{b}^{(1)}[k]+\mathbf{\Omega}_{b}^{(2)}[k]+\mathbf{\Omega}_{b}^{(3)}[k]+\underbrace{\sigma_\mr{N}^2\mathbf{W}_{\rf\mr{N}}^{H}\mathbf{W}_{\rf\mr{N}}}_\text{Gaussian noise}\notag\\&\overset{(a)}{=}\mathbf{\Omega}_{b}^{(1)}[k]+\mathbf{\Omega}_{b}^{(2)}[k]+\mathbf{\Omega}_{b}^{(3)}[k]+\sigma_\mr{N}^2\frac{n_R}{U}\mathbf{I}_U,
\end{align}
where $(a)$ is derived according to the property of RF beamformers, i.e.,  $\mathbf{W}_{\rf\mr{N}}^{H}\mathbf{W}_{\rf\mr{N}}=\frac{n_R}{U}\mathbf{I}_U$.
\begin{subequations}
\begin{align}
\mathbf{\Omega}_{b}^{(1)}[k]&=\mr{Cov}\bigg[\underbrace{\hat{\mathbf{H}}_\mr{ND}^{eff}[k]\mathbf{e}_\mr{D}[k]+\mathbf{\Delta}_\mr{ND}[k]\mathbf{e}_\mr{D}[k]}_{\text{backhaul channel transmitter HWI}}\notag\\&\quad+\underbrace{\mathbf{\Delta}_\mr{ND}[k]\mathbf{F}_{\bb\mr{D}}[k]\mathbf{s}_\mr{D}[k]}_{\text{backhaul channel estimation error}}\bigg]\notag\\&\overset{(b)}{=}\zeta\rho\hat{\mathbf{H}}_\mr{ND}^{eff}[k]\mr{diag}\left[\mathbf{F}_{\bb\mr{D}}[k]\mathbf{F}_{\bb\mr{D}}^H[k]\right]\left(\hat{\mathbf{H}}_\mr{ND}^{eff}[k]\right)^H\notag\\&\quad+\sigma_{e,\mr{ND}}^2\zeta(\rho+1)\mr{tr}\left[\mathbf{F}_{\bb\mr{D}}[k]\mathbf{F}_{\bb\mr{D}}^H[k]\right]\mathbf{I}_U,
\end{align}
where $(b)$ is obtained by following simplifications:
\begin{align}
    &\left[\mr{Cov}\left[\mathbf{\Delta}_\mr{ND}[k]\mathbf{e}_\mr{D}[k]\right]\right]_{m,n}\notag\\&=\sum_{p}\left[\mathbb{E}\left\{\left[\mathbf{\Delta}_\mr{ND}[k]\right]_{m,p}\left[\mathbf{\Delta}_\mr{ND}^H[k]\right]_{p,n}\left[\left\|\mathbf{e}_\mr{D}[k]\right\|^2\right]_{p}\right\}\right]_{m,n}\notag\\&=\sigma_{e,\mr{ND}}^2\sum_{p}\left[\mathbb{E}\left\{\left[\left\|\mathbf{e}_\mr{D}[k]\right\|^2\right]_{p}\right\}\right]_{m,n}\delta_{m,n}\notag\\&=\sigma_{e,\mr{ND}}^2\delta_{m,n}\mr{tr}\left[\mathbb{E}\left\{\mathbf{e}_\mr{D}[k]\mathbf{e}_\mr{D}^H[k]\right\}\right]\notag\\&=\sigma_{e,\mr{ND}}^2\zeta\rho\mr{tr}\left[\mr{diag}\left[\mathbf{F}_{\bb\mr{D}}[k]\mathbf{F}_{\bb\mr{D}}^H[k]\right]\right]\delta_{m,n}\notag\\&=\sigma_{e,\mr{ND}}^2\zeta\rho\mr{tr}\left[\mathbf{F}_{\bb\mr{D}}[k]\mathbf{F}_{\bb\mr{D}}^H[k]\right]\delta_{m,n},\label{est1}
\end{align}
\begin{align}
    &\left[\mr{Cov}\left[\mathbf{\Delta}_\mr{ND}[k]\mathbf{F}_{\bb\mr{D}}[k]\mathbf{s}_\mr{D}[k]\right]\right]_{m,n}\notag\\&=\zeta\sum_{p,q}\left[\mathbb{E}\left\{\left[\mathbf{\Delta}_\mr{ND}[k]\right]_{m,p}\left[\mathbf{F}_{\bb\mr{D}}[k]\mathbf{F}_{\bb\mr{D}}^H[k]\right]_{p,q}\left[\mathbf{\Delta}_\mr{ND}^H[k]\right]_{q,n}\right\}\right]_{m,n}\notag\\&=\sigma_{e,\mr{ND}}^2\zeta\sum_{p,q}\left[\mathbf{F}_{\bb\mr{D}}[k]\mathbf{F}_{\bb\mr{D}}^H[k]\right]_{p,q}\delta_{m,n}\delta_{p,q}\notag\\&=\sigma_{e,\mr{ND}}^2\zeta\mr{tr}\left[\mathbf{F}_{\bb\mr{D}}[k]\mathbf{F}_{\bb\mr{D}}^H[k]\right]\delta_{m,n}.\label{est2}
\end{align}
\end{subequations}
\begin{align}
\mathbf{\Omega}_{b}^{(2)}[n]&=\mr{Cov}\bigg[\underbrace{\hat{\mathbf{H}}_\mr{SI}^{eff}[k]\mathbf{e}_\mr{N}[k]+\mathbf{\Delta}_\mr{SI}[k]\mathbf{e}_\mr{N}[k]}_{\text{SI channel transmitter HWI}}\notag\\&\quad+\underbrace{\mathbf{\Delta}_\mr{SI}[k]\mathbf{F}_{\bb\mr{N}}[k]\mathbf{s}_\mr{N}[k]}_{\text{SI channel estimation error}}\bigg]\notag\\&\overset{(c)}{=}\zeta\rho\hat{\mathbf{H}}_\mr{SI}^{eff}[k]\mr{diag}\left[\mathbf{F}_{\bb\mr{N}}[k]\mathbf{F}_{\bb\mr{N}}^H[k]\right]\left(\hat{\mathbf{H}}_\mr{SI}^{eff}[k]\right)^H\notag\\&\quad+\sigma_{e,\mr{SI}}^2\zeta(\rho+1)\mr{tr}\left[\mathbf{F}_{\bb\mr{N}}[k]\mathbf{F}_{\bb\mr{N}}^H[k]\right]\mathbf{I}_U,
\end{align}
where $(c)$ is derived by using the similar simplification processes shown in \eqref{est1} and \eqref{est2}.
\begin{align}
    \mathbf{\Omega}_{b}^{(3)}[k]&=\underbrace{\beta\mr{diag}\left[\mr{Cov}\left[\widetilde{\hat{\mathbf{y}}}_\mr{N}[k]\right]\right]}_\text{receiver HWI}\notag\\&=\beta\mr{diag}\left[\mathbf{\Phi}_{b}[k]+\mathbf{\Omega}_{b}^{(1)}[k]+\mathbf{\Omega}_{b}^{(2)}[k]+\sigma_\mr{N}^2\frac{n_R}{U}\mathbf{I}_U\right].
\end{align}

Next, we will derive the sum SE expression of the access link across all users. The decoded signal at the $u$th user is given as
\begin{equation}
{y}_{\mr{E},u}[k]=\underbrace{\mathbf{w}_{\rf\mr{E},u}^H\left(\mathbf{H}_\mr{EN,u}[k]\mathbf{x}_\mr{N}[k]+\mathbf{z}_\mr{E,u}[k]\right)}_{\widetilde{{y}}_{\mr{E},u}[k]}+g_{\mr{E},u}[k].\label{yd2}
\end{equation}
By substituting \eqref{xd} and \eqref{RFA} into \eqref{yd2}, we can have the sum SE expression of the access link as follows, that is
\begin{equation}
\mathcal{R}_{a}=\sum_{u=1}^U\frac{1}{K}\sum_{k=1}^K\log_2\left(1+ \frac{{\Phi}_{a,u}[k]}{{\Omega}_{a,u}[k]}\right), \label{accesslink}
\end{equation}
where ${\Phi}_{a,u}[k]$ denotes the covariance for the known part of the $u$th user's desired signal and ${\Omega}_{a,u}[k]$ represents the covariance of the noise given by the multiuser interference, the channel estimation error, the transceiver HWI, and the Gaussian noise at the $u$th user.
\begin{equation}
{\Phi}_{a,u}[k]=\zeta\hat{\mathbf{h}}_{\mr{EN},u}^{eff}[k]\mathbf{f}_{\bb\mr{N},u}[k]\mathbf{f}_{\bb\mr{N},u}^H[k]\left(\hat{\mathbf{h}}_{\mr{EN},u}^{eff}[k]\right)^H,
\end{equation}
where $\hat{\mathbf{H}}_\mr{EN}^{eff}[k]=\left[\left(\hat{\mathbf{h}}_{\mr{EN},1}^{eff}[k]\right)^T,\left(\hat{\mathbf{h}}_{\mr{EN},2}^{eff}[k]\right)^T,\ldots,\left(\hat{\mathbf{h}}_{\mr{EN},U}^{eff}[k]\right)^T\right]^T$ with $\left\{\hat{\mathbf{h}}_{\mr{EN},u}^{eff}[k]\right\}_{u=1}^{U}\in\mathbb{C}^{1\times U}$.
\begin{align}
{\Omega}_{a,u}[k]&={\Omega}_{a,u}^{(1)}[k]+{\Omega}_{a,u}^{(2)}[k]+{\Omega}_{a,u}^{(3)}[k]+\underbrace{\sigma_\mr{E}^2\mathbf{w}_{\rf\mr{E},u}^H\mathbf{w}_{\rf\mr{E},u}}_\text{Gaussian noise}\notag\\&\overset{(d)}{=}{\Omega}_{a,u}^{(1)}[k]+{\Omega}_{a,u}^{(2)}[k]+{\Omega}_{a,u}^{(3)}[k]+\sigma_\mr{E}^2N_R,
\end{align}
where $(d)$ comes from the property of RF beamformers, i.e.,  $\mathbf{w}_{\rf\mr{E},u}^H\mathbf{w}_{\rf\mr{E},u}=N_R$.
\begin{align}
{\Omega}_{a,u}^{(1)}[k]&=\mr{Cov}\bigg[\underbrace{\sum_{v=1,v\neq u}^{U}\hat{\mathbf{h}}_{\mr{EN},u}^{eff}[k]\mathbf{f}_{\bb\mr{N},v}[k]{s}_{\mr{N},v}[k]}_{\text{multiuser interference}}+\underbrace{\hat{\mathbf{h}}_{\mr{EN},u}^{eff}[k]\mathbf{e}_{\mr{N}}[k]}_{\text{transmitter HWI}}\bigg]\notag\\&=\zeta\sum_{v=1,v\neq u}^{U}\hat{\mathbf{h}}_{\mr{EN},u}^{eff}[k]\mathbf{f}_{\bb\mr{N},v}[k]\mathbf{f}_{\bb\mr{N},v}^H[k]\left(\hat{\mathbf{h}}_{\mr{EN},u}^{eff}[k]\right)^H\notag\\&\quad+\zeta\rho\hat{\mathbf{h}}_{\mr{EN},u}^{eff}[k]\mr{diag}\left[\mathbf{F}_{\bb\mr{N}}[k]\mathbf{F}_{\bb\mr{N}}^H[k]\right]\left(\hat{\mathbf{h}}_{\mr{EN},u}^{eff}[k]\right)^H,
\end{align}
where $\mathbf{s}_\mr{N}[k]=\left[{s}_{\mr{N},1}[k],{s}_{\mr{N},2}[k],\ldots,{s}_{\mr{N},U}[k]\right]^T$.
\begin{align}
{\Omega}_{a,u}^{(2)}[k]&=\mr{Cov}\bigg[\underbrace{\mathbf{\Delta}_{\mr{EN},u}[k]\mathbf{e}_\mr{N}[k]}_{\text{transmitter HWI}}+\underbrace{\mathbf{\Delta}_{\mr{EN},u}[k]\mathbf{F}_{\bb\mr{N}}[k]\mathbf{s}_{\mr{N}}[k]}_{\text{channel estimation error}}\bigg]\notag\\&\overset{(e)}{=}\sigma_{e,\mr{EN}}^2\zeta(\rho+1)\mr{tr}\left[\mathbf{F}_{\bb\mr{N}}[k]\mathbf{F}_{\bb\mr{N}}^H[k]\right],
\end{align}
where $\mathbf{\Delta}_\mr{EN}[k]=\left[\left(\mathbf{\Delta}_{\mr{EN},1}[k]\right)^T,\left(\mathbf{\Delta}_{\mr{EN},2}[k]\right)^T,\ldots,\left(\mathbf{\Delta}_{\mr{EN},U}[k]\right)^T\right]^T$ with $\left\{\mathbf{\Delta}_{\mr{EN},u}[k]\right\}_{u=1}^{\mr{U}}\in\mathbb{C}^{1\times U}$ and $(e)$ is obtained by adopting the similar simplifications in \eqref{est1} and \eqref{est2}.
\begin{equation}
    {\Omega}_{a,u}^{(3)}[k]=\underbrace{\beta\left\|\widetilde{{y}}_{\mr{E},u}[k]\right\|^2}_\text{receiver HWI}=\beta\left({\Phi}_{a,u}[k]+{\Omega}_{a,u}^{(1)}[k]+{\Omega}_{a,u}^{(2)}[k]+\sigma_\mr{E}^2N_R\right).
\end{equation}
\subsection{Baseband Beamforming Design}
\label{hp}
Given the RF beamformers and RF effective channels derived from Section~\ref{RFest}, we aim to design the BB beamformers for both the backhaul and access links.

For the backhaul link, the $k$th BB precoder which maximizes the SE is obtained using the right singular vectors $\mathbf{V}_\mr{ND}[k]$ of the $k$th estimated RF effective backhaul link channel matrix $\hat{\mathbf{H}}^{eff}_\mr{ND}[k]$, that is
\begin{equation}
    \mathbf{F}_{\bb\mr{D}}[k]=\left[\mathbf{V}_\mr{ND}[k]\right]_{:,1:U}.
    \label{Fbb}
\end{equation}
Due to the precoder constraint, the BB precoder is updated as $\mathbf{F}_{\bb\mr{D}}[k]\leftarrow\frac{\sqrt{U}\mathbf{F}_{\bb\mr{D}}[k]}{\left|\left|\mathbf{F}_{\rf\mr{D}}\mathbf{F}_{\bb\mr{D}}[k]\right|\right|_F}$.

Next, the design of the BB precoder $\mathbf{F}_{\bb\mr{N}}[k]$ at the IAB-node transmitter aims to null the multiuser interference by the zero forcing, which is
\begin{equation}
   \mathbf{F}_{\bb\mr{N}}[k]=\left(\hat{\mathbf{H}}_\mr{EN}^{eff}[k]\right)^H\left[\hat{\mathbf{H}}_\mr{EN}^{eff}[k]\left(\hat{\mathbf{H}}_\mr{EN}^{eff}[k]\right)^H\right]^{-1}.
\end{equation}
Similarly, the BB precoder should be normalized as $\mathbf{f}_{\bb\mr{N},u}[k]\leftarrow\frac{\mathbf{f}_{\bb\mr{N},u}[k]}{\left|\left|\mathbf{F}_{\rf\mr{N}}\mathbf{f}_{\bb\mr{N},u}[k]\right|\right|_F}\;\forall u\in\{1,2,\ldots,U\}$.

Finally, with the fact that the channel estimation error is uncorrelated with the data vector, and we assume the strength of HWI and channel estimation error are known as a prior. The MMSE BB combiner for the $k$th subcarrier $\mathbf{W}_{\bb\mr{N}}[k]$ is designed by solving the following optimization problem, which is
\begin{equation}
   \arg\underset{\mathbf{W}_{\bb\mr{N}}[k]}\min{\mathbb{E}\left\{\left\|\mathbf{s}_\mr{D}[k]-\hat{\mathbf{y}}_\mr{N}[k]\right\|^2_2\right\}}.
\end{equation}
By solving $\frac{\partial\mathbb{E}\left\{\left\|\mathbf{s}_\mr{D}[k]-\hat{\mathbf{y}}_\mr{N}[k]\right\|^2_2\right\}}{\partial\mathbf{W}_{\bb\mr{N}}^H[k]}=0$ (see Appendix-\ref{a1}), we have
\begin{align}
   \mathbf{W}_{\bb\mr{N}}[k]&=\mathbb{E}\left\{\left({\widetilde{\hat{\mathbf{y}}}_\mr{N}[k]}+\mathbf{g}_\mr{N}[k]\right)\left({\widetilde{\hat{\mathbf{y}}}_\mr{N}[k]}+\mathbf{g}_\mr{N}[k]\right)^H\right\}^{-1}\notag\\&\quad\times\mathbb{E}\left\{\left({\widetilde{\hat{\mathbf{y}}}_\mr{N}[k]}+\mathbf{g}_\mr{N}[k]\right)\mathbf{s}_\mr{D}^H[k]\right\}\notag\\&=\zeta\left(\mathbf{\Phi}_{b}[k]+\mathbf{\Omega}_{b}[k]\right)^{-1}\hat{\mathbf{H}}_\mr{ND}^{eff}[k]\mathbf{F}_{\bb\mr{D}}[k].\label{MMSEBB}
\end{align}

\begin{table*}[!t]
\renewcommand{\arraystretch}{0.9}
\caption{System Parameters and Default Values}
\label{notation}
\centering
\begin{threeparttable}
\begin{tabular}{|c|c|c|}
\hline
\textbf{Notation} & \textbf{Physical Meaning} & \textbf{Values}\\
\hline
$K$ & Number of subcarriers & 512\\
\hline
$D$ & Number of cyclic prefixes & 128\\
\hline
$W$ & Bandwidth & 400 MHz\\
\hline
$f_c$ & Carrier frequency & 28 GHz\\
\hline
$U$ & Number of users (subarrays, RF chains, data streams) & 4\\
\hline
${N_T},{n_T}$ & Number of transmit antennas at the IAB donor and the IAB-node, respectively & $16\times16$, $16\times16$\\
\hline
${n_R},{N_R}$ & Number of receive antennas at the IAB-node and each user, respectively & $16\times16$, $16\times4$\\
\hline
$\sigma_\mr{N},\sigma_\mr{E}$ & Gaussian noise power at the IAB-node and each user, respectively & 
\tabincell{l}{$-174\;\text{dBm}+10\log_{10}W$\\$+10$ dB}\\
\hline
$T_s$ & Sampling time & $1/W$\\
\hline
$\tau_{c,l}$ & Path delay & $\mathcal{U}(0, DT_s)$\\ 
\hline
$\alpha_{c,l}$ & Complex gain & $\mathcal{CN}(0,1)$\\
\hline
$r_0$ & Reference distance & 1 m\\
\hline
$r$ & Distance between transceiver, respectively & 100 m (0.1 m)\tnote{4}\\
\hline
$\mu$ & Path loss exponent & 3.4 \cite{7414036}\\
\hline
$\lambda$ & Wavelength & $3\times10^8/f_c$\\
\hline
$\kappa$ & Rician factor & 10 dB\\
\hline
$\eta$ & Power of the SI signal attenuated by antenna isolation and A-SIC & -80 dB\\
\hline
\end{tabular}
\begin{tablenotes}
        \footnotesize
        \item[4] $r=100$ m for backhaul and access link channels; $r=0.1$ m ($\approx10\lambda$ \cite{9013116}) for SI channel.
\end{tablenotes}
\end{threeparttable}
\end{table*}
\section{Simulations}\label{sec5}
In this section, simulation results will be shown to analyze the performance of our designed networks. Each subarray (users) has 16$\times$4 UPA with 1 RF chain. The rolling factor of the pulse shaping filter is 1. Both communication links have $\mr{N_C}=8$ clusters, each with $\mr{N_L}=10$ rays, whereas the NLOS component of the SI channel has $\mr{N_C}=2$ clusters, each with $\mr{N_L}=8$ rays. Both azimuth and elevation AOAs/AODs can be expressed as the sum of the mean angle of each cluster and the angle shifts in the cluster. The mean azimuth and elevation AOAs/AODs of each cluster are assumed as uniformly distributed in $[-\pi,\pi]$, and $\left[-\frac{\pi}{2},\frac{\pi}{2}\right]$, respectively. In each cluster, the AOAs/AODs have Laplacian distribution with an angle spread of $5^\circ$. The transceiver arrays at the IBFD-IAB-node have a separation angle of $\frac{\pi}{6}$. Assume $\sigma^2_\mr{N}=\sigma^2_\mr{E}=\sigma^2$, we define $\mr{SNR}\triangleq\frac{P_\mr{r}}{\sigma^2KU}$, where $P_\mr{r}=\frac{P_\mr{t}}{\bar{PL}}$ is the ratio between transmit power and average path loss according to the Friis’ law. We let HWI factors $\rho=\beta$ be the same for all channels. The backhaul link SE of the HD scheme is given by removing the part relevant to the SI in \eqref{backhaullink} due to non-simultaneous transmission and reception. Moreover, for both links, the (sum) SE expressions for HD transmission need to be scaled by 0.5 since separate time-frequency signaling channels are used for backhaul and access link. Other parameters and their default values used in the simulations are summarized in Table~\ref{notation}.

\begin{figure}[t!]
\centering
\subfigure[]{
\includegraphics[width=0.45\textwidth]{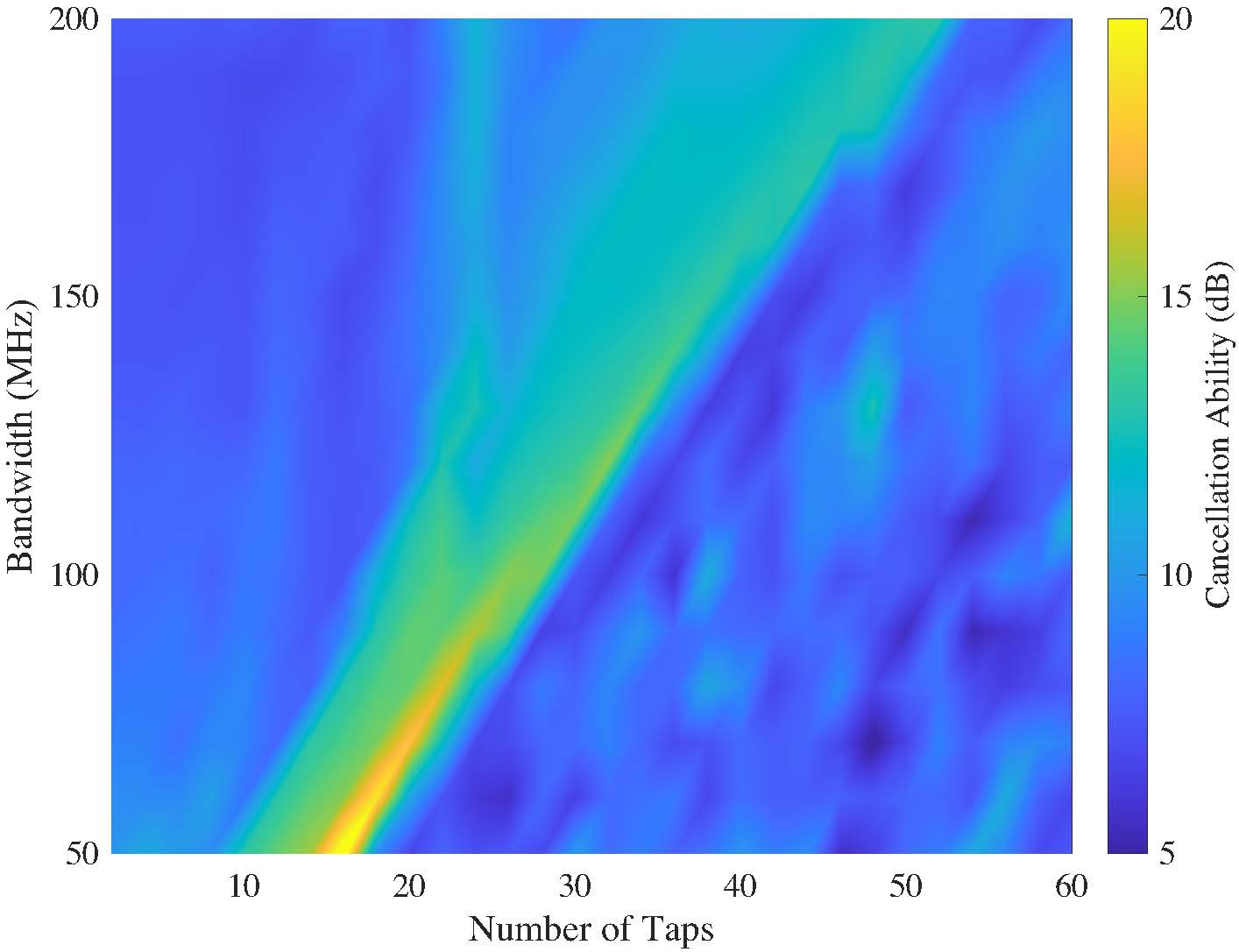}\label{xx}}
\subfigure[]{
\includegraphics[width=0.45\textwidth]{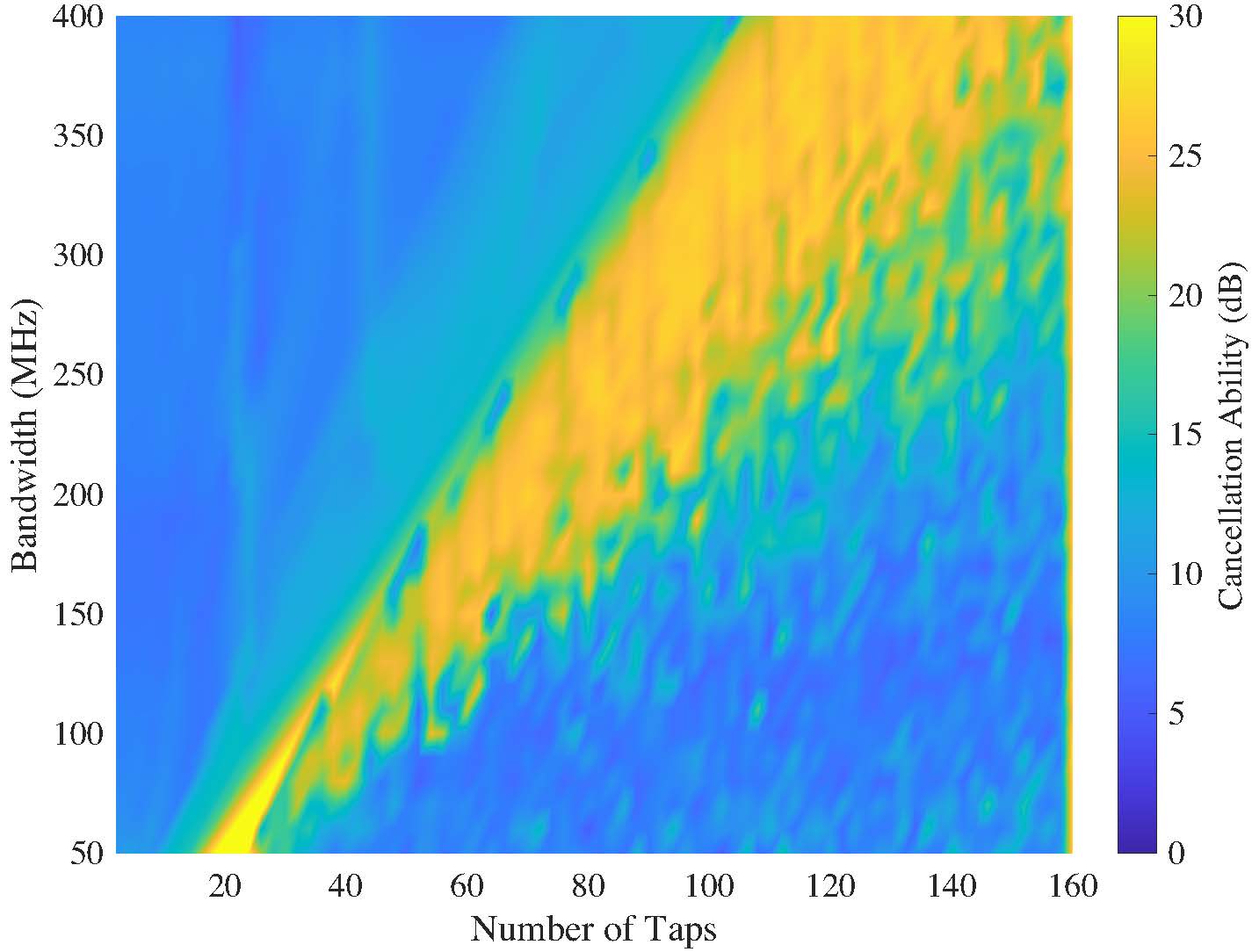}\label{yy}}
\caption{Comparison between the performance of (a) traditional micro-strip analog canceler; (b) OD-based analog canceler (SI channel has a delay spread of 200 ns).}
\label{ASIC}
\end{figure}
\subsection{Performance of OD-Based Analog Canceler}
Assume the propagation loss of the FBG (coiled into 2 cm) is 0.461 dB/m, and that of the micro-strip is 2.967 dB/m \cite{8796422}. The OD-based design uses a 20 dB hybrid coupler to couple the RF reference signal into the OD-based canceler, while the conventional  electrical canceler uses a 0 dB coupler. Besides, to explore the best performance, the tap delay varies according to the number of taps to cover the delay spread. Fig.~\ref{ASIC} shows the A-SIC abilities (in dB) of the traditional micro-strip canceler (see Fig.~\ref{xx}) and the OD-based canceler (see Fig.~\ref{yy}) for different bandwidths and numbers of taps. Simulations are run with 200 ns of significant delay spread for the SI channel, which reflects a bad channel condition. Although a measurement for the SI channel delay spread is done in \cite{7414127}, a general delay spread value is still lacking in the literature. Fig.~\ref{xx} shows that creating a large number of taps with conventional electrical components (e.g., cables or micro-strips) degrades the performance rather than improving it due to significant insertion losses. It can be seen that less than 15 dB of cancellation is achieved under 200 MHz bandwidth. Fig.~\ref{yy} shows that under 400 MHz bandwidth, OD-based canceler can achieve around 25 dB of cancellation in FR2 wideband with 100 taps, which is also proved in \cite{8796422}. Note that this result shows the cancellation ability that can be achieved between a single RF chain pair. In this work, we assume antenna isolation and our A-SIC can attenuate the SI signal power by 55 dB \cite{6702851} and 25 dB, respectively.

\begin{figure}[t!]
\centering
\subfigure[]{
\includegraphics[width=0.45\textwidth]{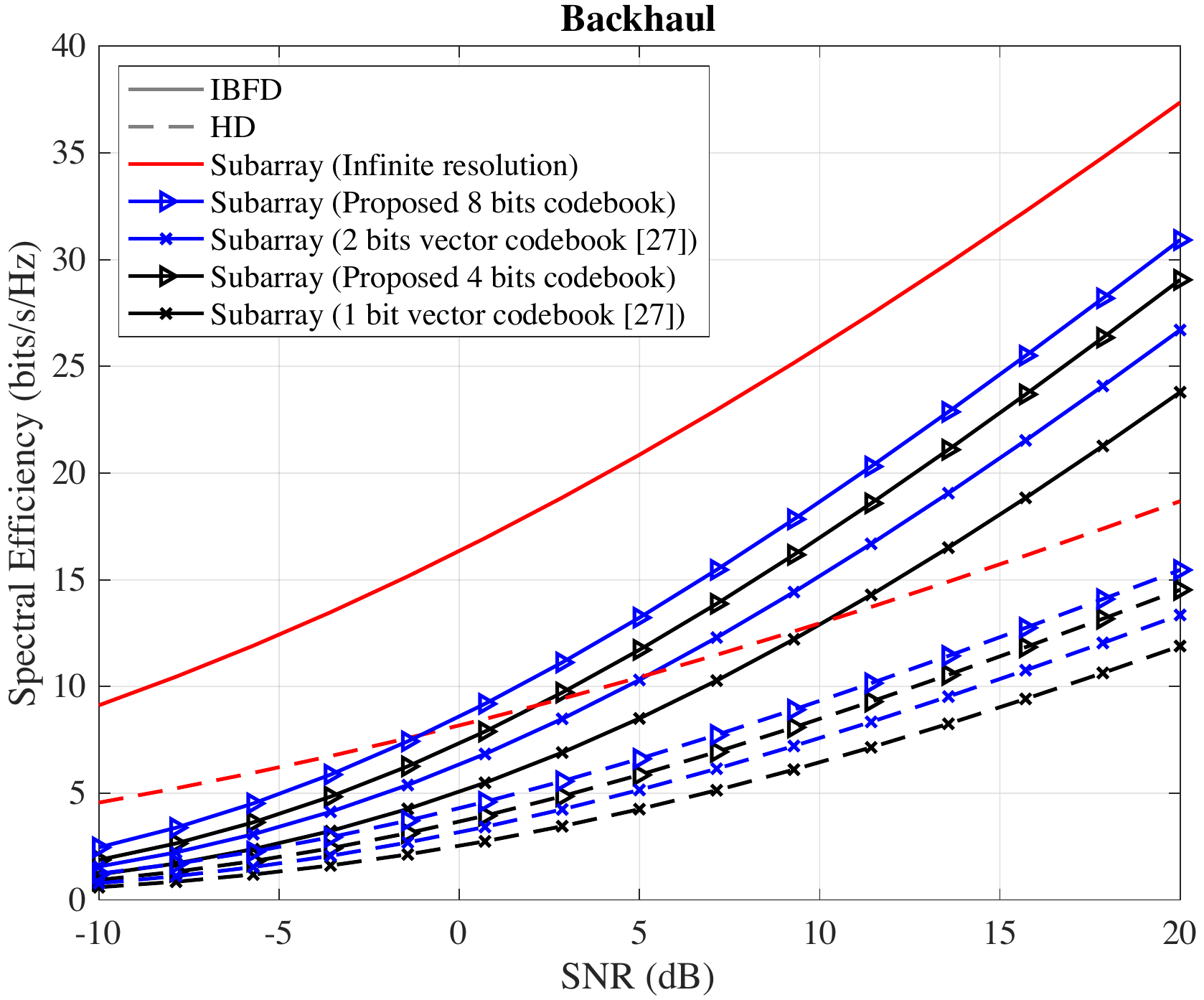}}
\subfigure[]{
\includegraphics[width=0.45\textwidth]{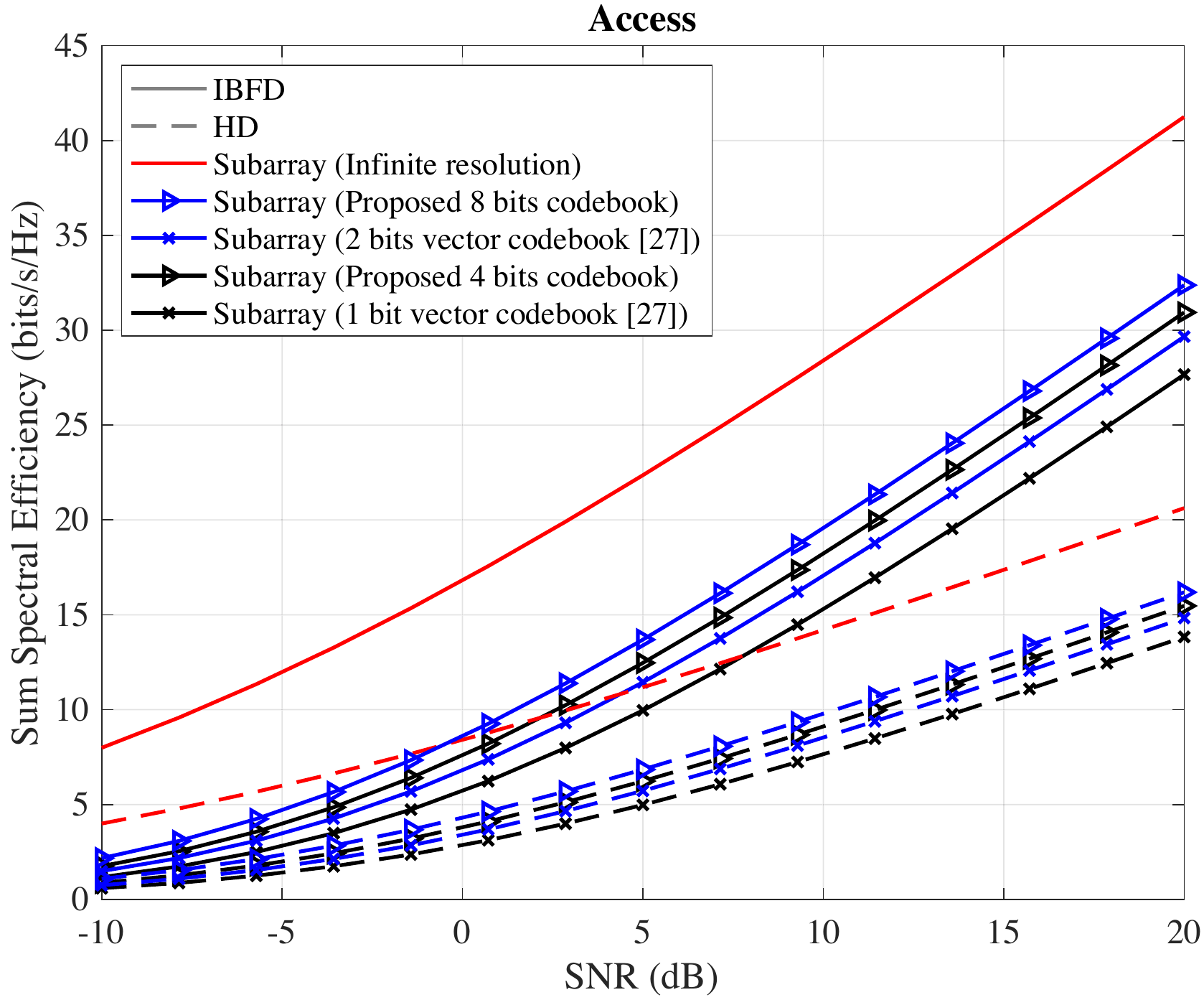}}
\caption{Comparison on the (sum) SE of 4-subarray hybrid beamforming structure with different kinds of codebooks for (a) backhaul link; (b) access link with 4 users. Each
subarray (user) is equipped with $16\times4$ UPA and 1 RF chain (perfect CSI without HWI).}
\label{Quan}
\end{figure}
\subsection{Performance of the Proposed Codebook Design}
The comparison on the (sum) SE of the backhaul and access links with RF precoders/combiners selected from our proposed matrix-wise codebooks and vector-wise codebooks designed by conventional MSE-based LBG algorithm in \cite{1094577}, respectively, for the subarray structure is plotted in Fig.~\ref{Quan}. In order to get a fair comparison, a $b$-bit vector codebook should be compared with an $N_{\rf}b$-bit matrix codebook, where $N_\rf$ is the number of RF chains\footnote[5]{For vector quantization, since each column of the RF beamformer matrix selects one codeword from a $b$-bit vector codebook, we can get $2^{N_{\rf}b}$ different candidate matrices, which is equal to the number of codeworks in a $N_{\rf}b$-bit matrix codebook.}. We assume perfect CSI and hardware. The RF precoders/combiners with infinite resolution PSs are designed according to \cite{7880698}. It can be seen that, for both kinds of codebooks, as the number of codebook size increases, the performance becomes closer to the ideal one (i.e., infinite resolution). Obviously, our matrix codebook can provide better performance than the vector codebook designed by the conventional LBG algorithm, which shows the successful applicability of our modified MSE-based LBG codebook design. However, there is still a small gap between the ideal one and the curves derived with 8 bits matrix codebook for both links. A large size of codebook can be used to reduce the gap. Moreover, the HD operation yields lower (sum) SE than that of the IBFD scheme.

\begin{figure}
\centering
\includegraphics[width=0.45\textwidth]{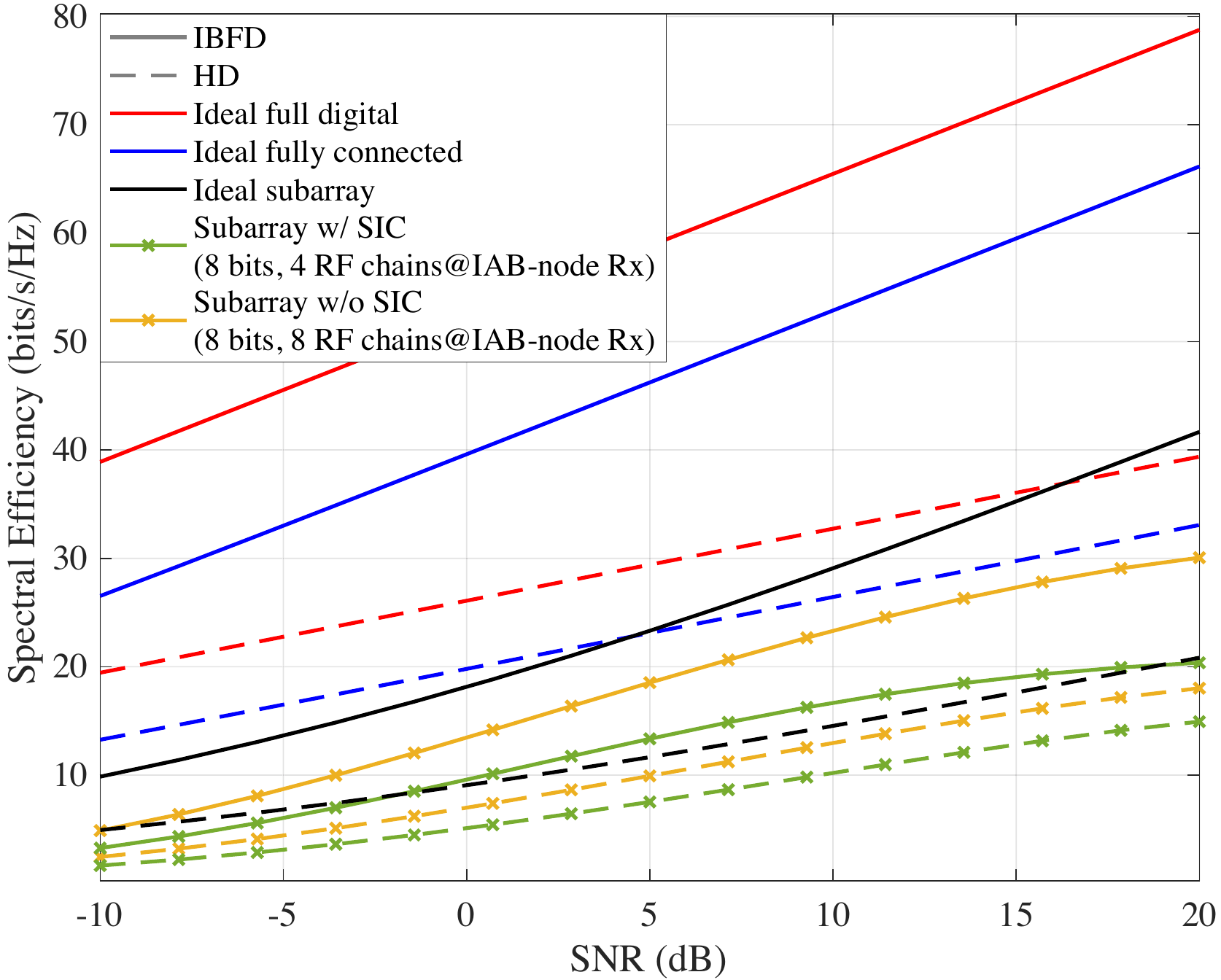}
\caption{SE of backhaul link for different beamforming schemes. Each subarray (user) has $16\times4$ UPA. The IAB donor and IAB-node have $16\times16$ UPA for fully connected structure. ($\rho=\beta=-80$ dB, $\sigma_{e,\mr{ND}}^2=\sigma_{e,\mr{EN}}^2=\sigma_{e,\mr{SI}}^2=-120$ dB)}\label{SIII}
\end{figure}
\subsection{Performance of Different Beamforming Schemes}
Fig.~\ref{SIII} shows the SE of the backhaul link for different beamforming schemes. The ideal curves are plotted by assuming perfect CSI and SIC without HWI. The design of the RF precoders/combiners for the ideal fully connected and subarray structures follows the process in \cite{7880698}, which have infinite resolution. The non-ideal curves are plotted by our proposed design algorithm with 8 bits RF codebook and setting $\rho=\beta=-80$ dB, $\sigma_{e,\mr{ND}}^2=\sigma_{e,\mr{EN}}^2=\sigma_{e,\mr{SI}}^2=-120$ dB. It can be observed that for the IBFD scheme, these three beamforming schemes evaluated in the figure are separated by a significant rate loss. Although the rate loss is evident, the subarray structure can significantly reduce the hardware complexity and provide low-computationally intensive precoders, which is beneficial for industrial implementations. Further, with our staged SIC, the SE of the subarray structure is very close to its ideal one; however, it shows some degrees of freedom loss at high SNR due to RSI caused by HWI and RF effective channel uncertainties. Fortunately, the losses on degrees of freedom and SE are further reduced by increasing the number of RF chains at the IAB-node receiver from 4 to 8 (see the green and orange curves in Fig.~\ref{SIII}). 

\begin{figure}[t!]
\centering
\subfigure[]{
\includegraphics[width=0.45\textwidth]{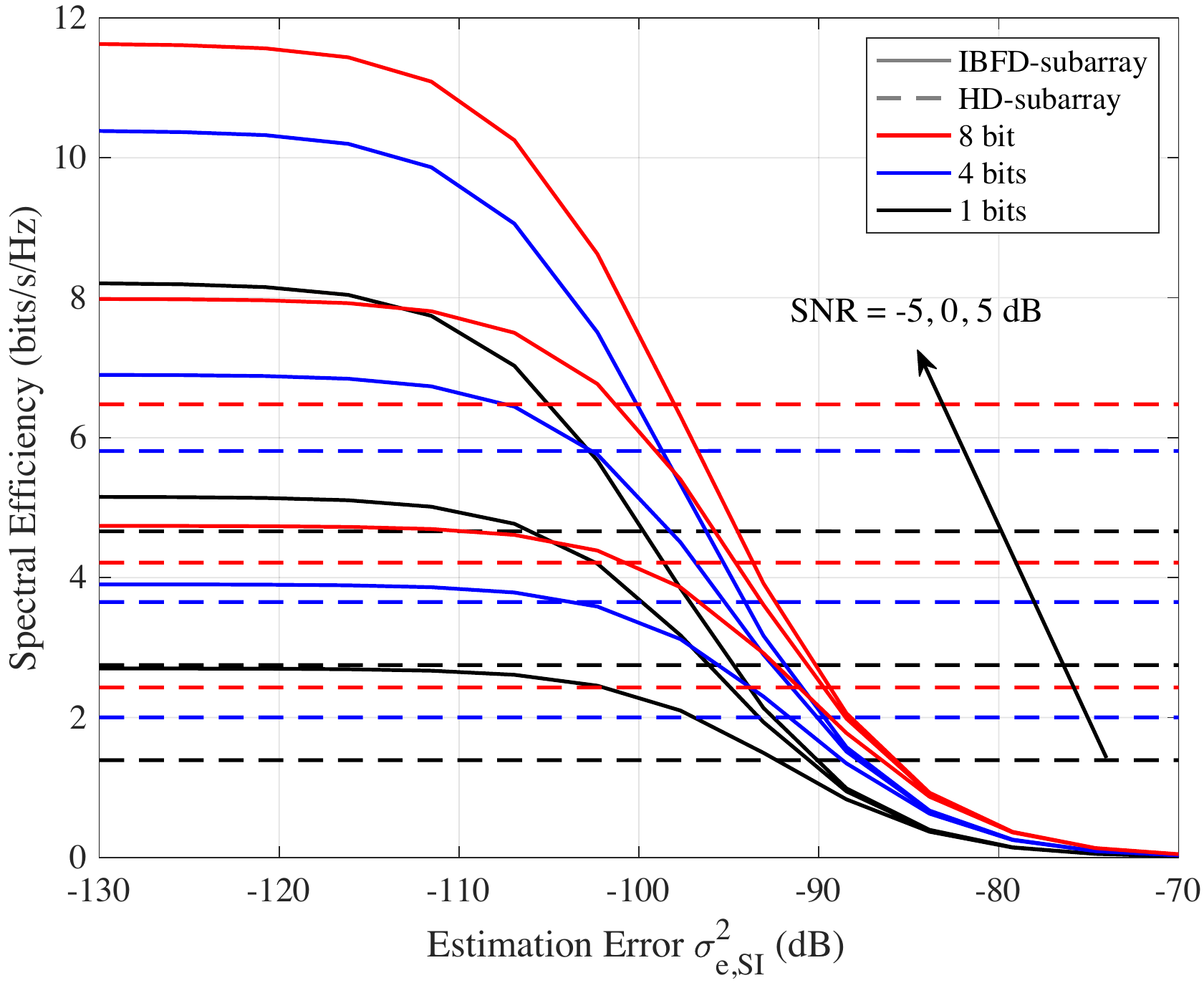}\label{CEER}}
\subfigure[]{
\includegraphics[width=0.45\textwidth]{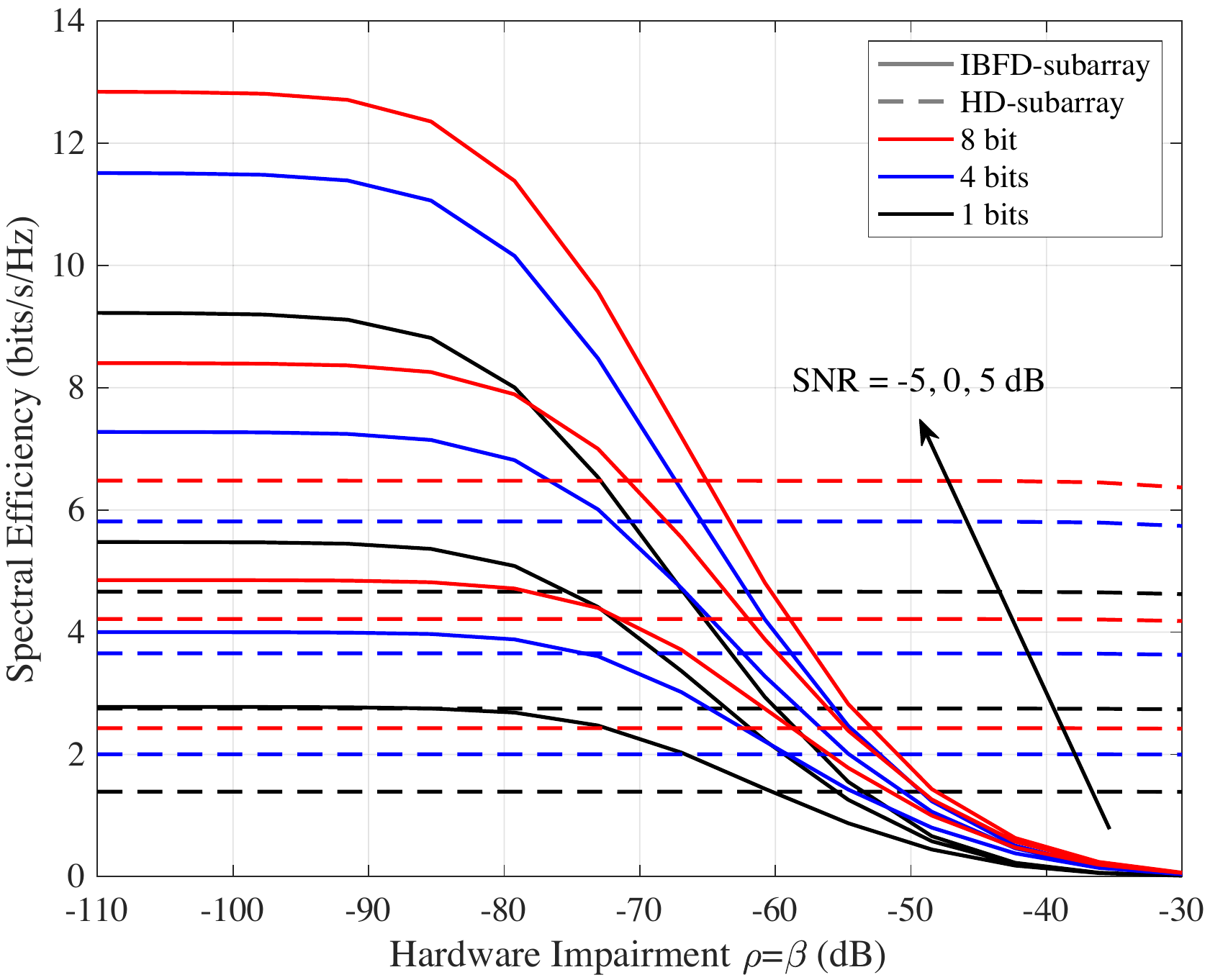}\label{HWIR}}
\caption{SE of the backhaul link at $\text{SNR}=-5,0,5$ dB with 4-subarray hybrid beamforming structure, where RF beamformers are selected from different size of codebooks, in the presence of different values of (a) SI RF effective channel estimation error ($\rho=\beta=-80$ dB, $\sigma_{e,\mr{ND}}^2=\sigma_{e,\mr{EN}}^2=-120$ dB); (b) HWI ($\rho=\beta$, $\sigma_{e,\mr{ND}}^2=\sigma_{e,\mr{EN}}^2=\sigma_{e,\mr{SI}}^2=-120$ dB).}
\label{HWICEE}
\end{figure}
\subsection{Effect of RSI on the SE of the Backhaul Link}
In Fig.~\ref{HWICEE}, with RF precoders/combiners selected from 1, 4, 8 bits codebooks, respectively, we would like to study how the RSI caused by RF effective SI channel estimation error and HWI can affect the SE performance of the backhaul link at different SNR values. With $\rho=\beta=-80$ dB and $\sigma_{e,\mr{ND}}^2=\sigma_{e,\mr{EN}}^2=-120$ dB, we plot the SE performance of the backhaul link in Fig.~\ref{CEER} by varying the channel estimation error of the SI RF effective channel. Interestingly, it is worth noting that as the size of the RF codebook increases, the intersection point (i.e., the point where both the IBFD and HD have the same performance) shifts to the right at a fixed SNR, which means the system can tolerate more RSI caused by channel estimation error. On the contrary, when the codebook size is fixed, as SNR increases, the intersection point shifts to the left. By assuming all effective channels have the same estimation error of -120 dB, Fig.~\ref{HWIR} shows the backhaul link SE performance with varying HWI factors. Similar to the trend in Fig.~\ref{CEER}, with the same codebook size, as the SNR increases, the system can tolerate less RSI caused by HWI. The tolerance is improved at a fixed SNR when the codebook size increases. Moreover, an almost doubled SE can be achieved by the IBFD compared to that of the HD when HWI factors (and channel estimation errors) are small enough, as can be seen in Fig.~\ref{HWICEE}.

\section{Conclusion}\label{sec6}
In this paper, we have studied FR2 wideband IBFD-IAB networks under subarray structures, which are simpler to deploy and more cost-effective than fully-connected ones. For this system, we have proposed the RF codebook design for the subarray structure with hybrid beamforming. Compared with the traditional vector-wise codebook, our matrix-wise codebook can avoid low-rank matrix and loss of degrees of freedom. We also introduced the staged SIC scheme. In order to reduce the deployment cost, we have established the canceler on each RF chain pair and utilized the OD-based analog canceler to reduce the effect of insertion loss. The RSI left by the A-SIC was handled in the digital domain by successive interference cancellation and MMSE BB combiner. Simulations have shown that under 400 MHz bandwidth, our OD-based canceler can achieve about 25 dB cancellation with 100 taps as well as experiencing constant insertion loss, which can not be realized by the traditional micro-strip canceler. With large HWI and RF effective SI channel uncertainties, the IBFD transmission experiences performance limitation in the backhaul link; however, for small HWI and uncertainties, the IBFD promises almost doubled SE compared with that of the HD.

Further work will include investigating multicell IBFD-IAB systems, optimal power allocation, and efficient antenna cancellation. Besides, the SI channel model will also be studied by real-world measurements or other reliable mathematics models.

\appendices
\section{MMSE BB Combiner}\label{a1}
\begin{align}
    &\frac{\partial\mathbb{E}\left\{\left\|\mathbf{s}_\mr{D}[k]-\hat{\mathbf{y}}_\mr{N}[k]\right\|^2_2\right\}}{\partial\mathbf{W}_{\bb\mr{N}}^H[k]}\notag\\&=\frac{\partial\mathbb{E}\left\{(\mathbf{s}_\mr{D}[k]-\hat{\mathbf{y}}_\mr{N}[k])(\mathbf{s}_\mr{D}[k]-\hat{\mathbf{y}}_\mr{N}[k])^H\right\}}{\partial\mathbf{W}_{\bb\mr{N}}^H[k]}\notag\\&=\frac{\partial\mathbb{E}\left\{\mathbf{s}_\mr{D}[k]\mathbf{s}^H_\mr{D}[k]-\mathbf{s}_\mr{D}[k]\hat{\mathbf{y}}_\mr{N}^H[k]-\hat{\mathbf{y}}_\mr{N}[k]\mathbf{s}^H_\mr{D}[k]+\hat{\mathbf{y}}_\mr{N}[k]\hat{\mathbf{y}}_\mr{N}^H[k]\right\}}{\partial\mathbf{W}_{\bb\mr{N}}^H[k]}.\label{43}
\end{align}

By substituting \eqref{RFres} into \eqref{43}, we have
\begin{align}
    &\frac{\partial\mathbb{E}\left\{\left\|\mathbf{s}_\mr{D}[k]-\hat{\mathbf{y}}_\mr{N}[k]\right\|^2_2\right\}}{\partial\mathbf{W}_{\bb\mr{N}}^H[k]}=-\left({\widetilde{\hat{\mathbf{y}}}_\mr{N}[k]}+\mathbf{g}_\mr{N}[k]\right)\mathbf{s}^H_\mr{D}[k]\notag\\&+\left({\widetilde{\hat{\mathbf{y}}}_\mr{N}[k]}+\mathbf{g}_\mr{N}[k]\right)\left({\widetilde{\hat{\mathbf{y}}}_\mr{N}[k]}+\mathbf{g}_\mr{N}[k]\right)^H\mathbf{W}_{\bb\mr{N}}[k].\label{44}
\end{align}

Let \eqref{44} equal to 0, we can have the MMSE BB combiner given in \eqref{MMSEBB}.

\ifCLASSOPTIONcaptionsoff
  \newpage
\fi
\bibliographystyle{IEEEtran}
\bibliography{IEEEabrv,Ref}

\begin{IEEEbiography}
[{\includegraphics[width=1in,height=1.25in,clip,keepaspectratio]{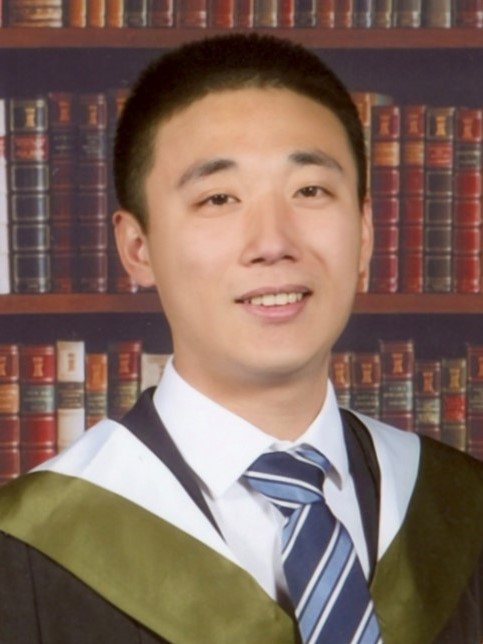}}]{Junkai Zhang}
(Student Member, IEEE) received the B.Eng. degree in communication engineering from Shenyang Ligong University, Shenyang, China, in 2018, and the M.Sc. degree in 2019, in signal processing and communications (with Distinction) from The University of Edinburgh, Edinburgh, United Kingdom, where he is currently working toward the Ph.D. degree with the Institute for Digital Communications. His research interests include 5G and beyond wireless networks, millimeter-wave communications, in-band-full-duplex radio, stochastic geometry, integrated sensing and communications, and massive MIMO.
\end{IEEEbiography}
\vfill
\begin{IEEEbiography}
[{\includegraphics[width=1in,height=1.25in,clip,keepaspectratio]{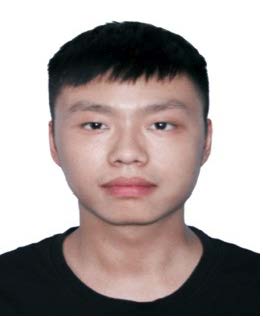}}]{Haifeng Luo}
received the Bachelor’s degree from the Civil Aviation University of China, Tianjin, China, in 2018, and received the Master's degree from The University of Edinburgh, Edinburgh, United Kingdom, in 2019. He is currently pursuing the Ph.D. degree in the Institute for Digital Communications, The University of Edinburgh, Edinburgh, United Kingdom. His research interests include signal processing aspects of beyond 5G wireless networks, full-duplex radios, and machine learning aided communications.
\end{IEEEbiography}
\vfill
\begin{IEEEbiography}
[{\includegraphics[width=1in,height=1.25in,clip,keepaspectratio]{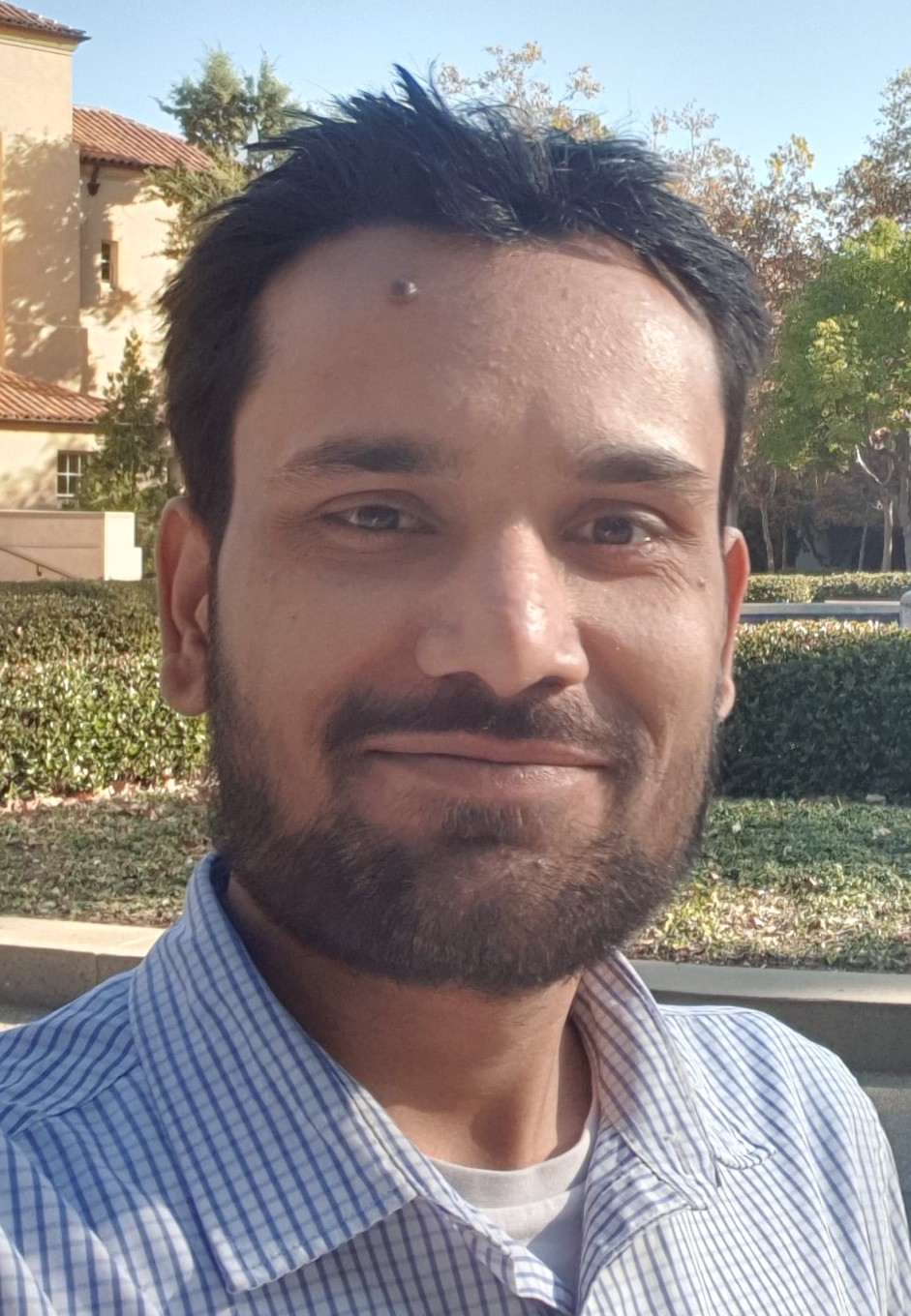}}]{Navneet Garg}
(Member, IEEE) received the B.Tech. degree in electronics and communication engineering from College of Science \& Engineering, Jhansi, India, in 2010, and the M.Tech. degree in digital communications from ABV-Indian Institute of Information Technology and Management, Gwalior, in 2012. He has completed the Ph.D. degree in June 2018 from the department of electrical engineering at the Indian Institute of Technology Kanpur, India. From July 2018-Jan 2019, he visited The University of Edinburgh, Edinburgh, United Kingdom. From February 2019-2020, he is employed as a research associate in Heriot-Watt university, Edinburgh, United Kingdom. Since February 2020, he is working as a research associate in The University of Edinburgh, Edinburgh, United Kingdom. His main research interests include wireless communications, signal processing, optimization, and machine learning.
\end{IEEEbiography}
\vfill
\begin{IEEEbiography}
[{\includegraphics[width=1in,height=1.25in,clip,keepaspectratio]{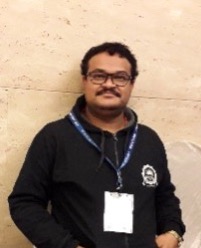}}]{Abhijeet Bishnu}
(Member, IEEE) received his B.Eng. degree in Electronics and Communication Engineering from Technocrat Institute of Technology, Bhopal, India in 2010. He also received the M.Eng. degree in Electronics and Telecommunication Engineering from S.G.S.I.T.S. Indore, India in 2013, and the Ph.D. degree in Electrical Engineering from Indian Institute of Technology Indore, India, in 2019. He was also a visiting research scholar in The University of Edinburgh, Edinburgh, United Kingdom in 2019. He is currently a postdoctoral research associate in The University of Edinburgh, Edinburgh, United Kingdom. His research interests include channel estimation, cognitive radio, MIMO-OFDM system, full-duplex communication and 5G and beyond communication. He is served as reviewer for many IEEE and Springer journals.
\end{IEEEbiography}
\vfill
\begin{IEEEbiography}
[{\includegraphics[width=1in,height=1.25in,clip,keepaspectratio]{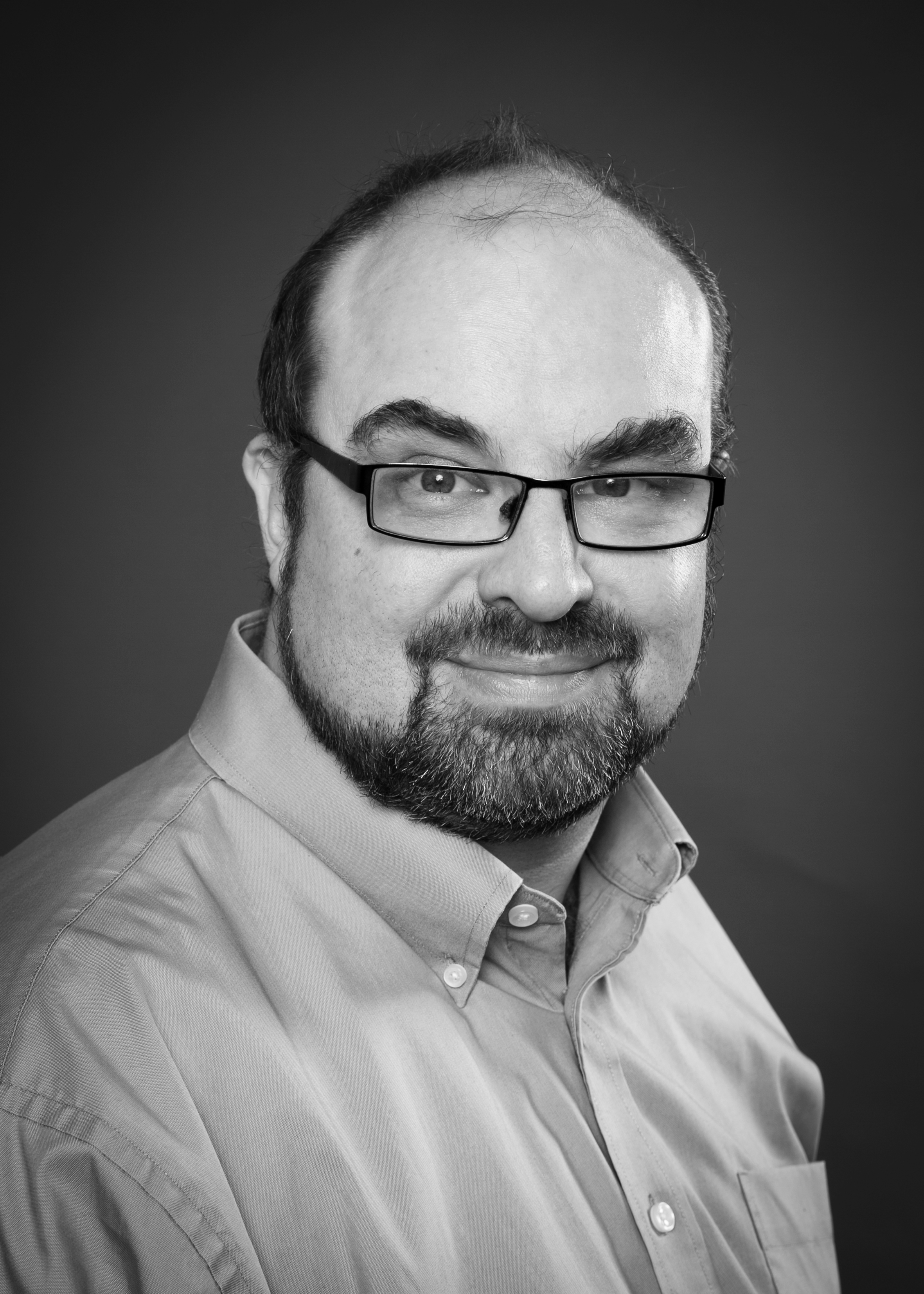}}]{Mark Holm}
(Member, IEEE) received the B.S. degree (Hons.) in laser physics and optoelectronics and the Ph.D. degree in physics from the University of Strathclyde, in 1997 and 2001, respectively. He currently works as the Technical Lead and a Hardware System Architect with Huawei Technologies (Sweden) AB, with interest in microwave radio, phased array antennas, full duplex radio systems, and photonic radios. In the past, he was the Microwave Lead on AESA radar systems, a Senior Engineer responsible for GaAs pHemt modeling, and a Laser and Package Design Engineer for SFP/XENPACK fiber modules. He has published in the fields of laser design and GaAs device modeling.
\end{IEEEbiography}
\vfill
\begin{IEEEbiography}
[{\includegraphics[width=1in,height=1.25in,clip,keepaspectratio]{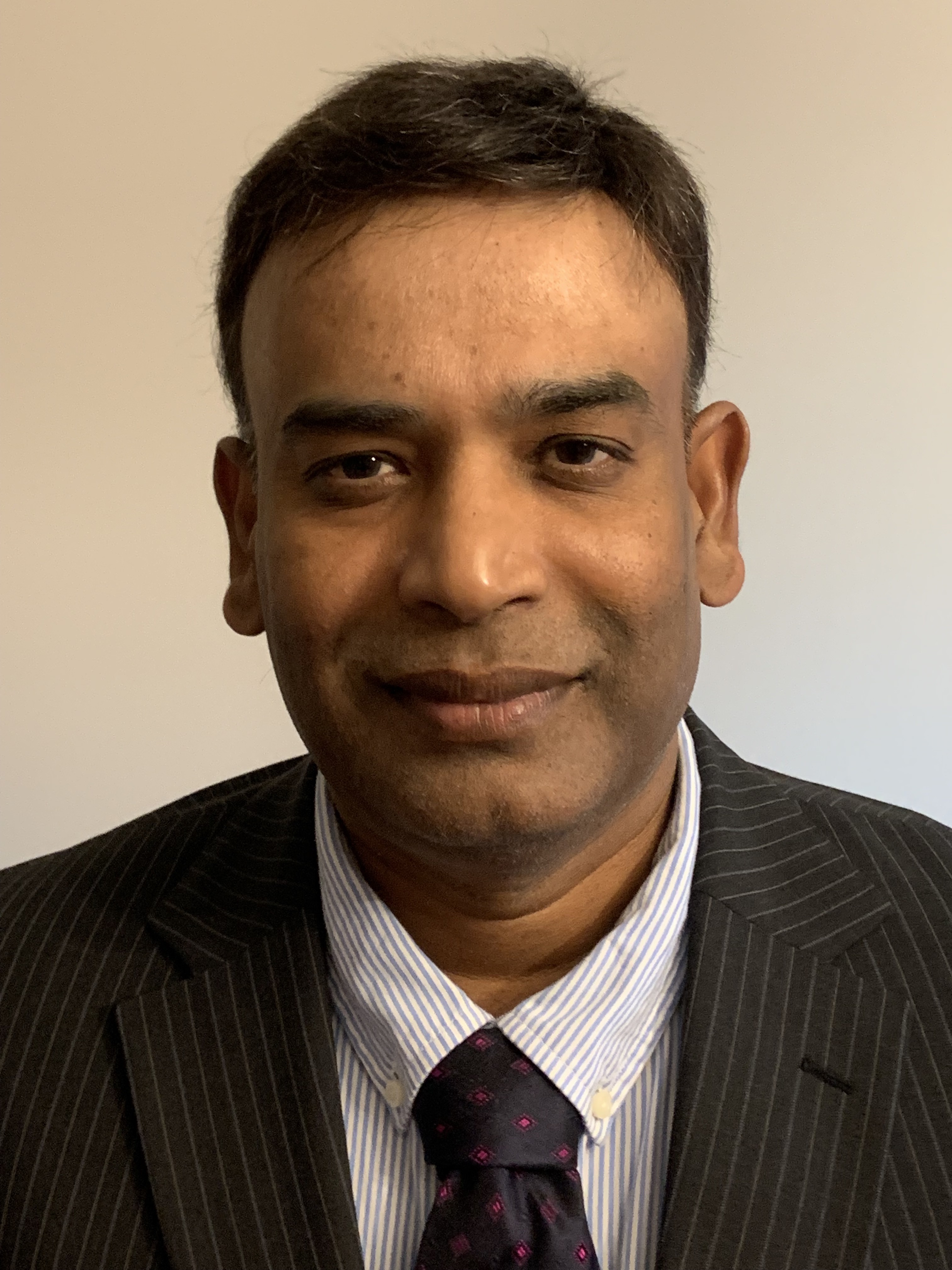}}]{Tharmalingam Ratnarajah}
(Senior Member, IEEE) is currently with the Institute for Digital Communications, The University of Edinburgh, Edinburgh, UK, as a Professor in Digital Communications and Signal Processing. He was a Head of the Institute for Digital Communications during 2016-2018. His research interests include signal processing and information theoretic aspects of beyond 5G wireless networks, full-duplex radio, mmWave communications, random matrices theory, interference alignment, statistical and array signal processing and quantum information theory. He has published over 400 publications in these areas and holds four U.S. patents. He has supervised 16 PhD students and 21 post-doctoral research fellows and raised $\$$11+ million USD of research funding. He was the coordinator of the EU projects ADEL (3.7M \EUR) in the area of licensed shared access for 5G wireless networks, HARP (4.6M \EUR) in the area of highly distributed MIMO, as well as EU Future and Emerging Technologies projects HIATUS (3.6M \EUR) in the area of interference alignment and CROWN (3.4M \EUR) in the area of cognitive radio networks. Dr Ratnarajah was an associate editor IEEE Transactions on Signal Processing, 2015-2017 and Technical co-chair, The 17th IEEE International workshop on Signal Processing advances in Wireless Communications, Edinburgh, UK, 3-6, July 2016. Dr Ratnarajah is a Fellow of Higher Education Academy (FHEA).
\end{IEEEbiography}
\end{document}